\documentclass[]{spie}  %>>> use for US letter paper
\usepackage{amsmath,amsfonts,amssymb}
\usepackage{graphicx}
\usepackage[colorlinks=true, allcolors=blue]{hyperref}

\title{SPIE Proceedings: Two decades of Exoplanetary Science with Adaptive Optics}

\author[a,b]{Chauvin G.}
\affil[a]{Unidad Mixta Internacional Franco-Chilena de Astronom\'{i}a, CNRS/INSU UMI 3386 and Departamento de Astronom\'{i}a, Universidad de Chile, Casilla 36-D, Santiago, Chile}
\affil[b]{Univ. Grenoble Alpes, CNRS, IPAG, F-38000 Grenoble, France}

\authorinfo{Further author information: E-mail: gael.chauvin@univ-grenoble-alpes.fr}

\begin{document} 
\maketitle

\begin{abstract}

 As astronomers, we are living an exciting time for what concerns the search for other worlds.  Recent discoveries have already deeply impacted our vision of planetary formation and architectures. Future bio-signature discoveries will probably deeply impact our scientific and philosophical understanding of life formation and evolution. In that unique perspective, the role of observation is crucial to extend our understanding of the formation and physics of giant planets shaping planetary systems. With the development of high contrast imaging techniques and instruments over more than two decades, vast efforts have been devoted to detect and characterize lighter, cooler and closer companions to nearby stars, and ultimately image new planetary systems. Complementary to other planet-hunting techniques, this approach has opened a new astrophysical window to study the physical properties and the formation mechanisms of brown dwarfs and planets.  I will briefly review the different observing techniques and strategies used, the main samples of targeted stars, the key discoveries and surveys, to finally address the main results obtained so far about the physics and the mechanisms of formation and evolution of young giant planets and planetary system architectures.   
\end{abstract}


%%%%%%%%%%%%%%%%%%%%%%%%%%%%%%%%%%%%%%%%%%%%%%%%%%%%%%%%%%%%%%%%%%%%%%%%%%%%%
\section{Introduction}
\label{sec:intro}  % \label{} allows reference to this section

\begin{figure}[t]
\begin{center}
\includegraphics[height=5cm]{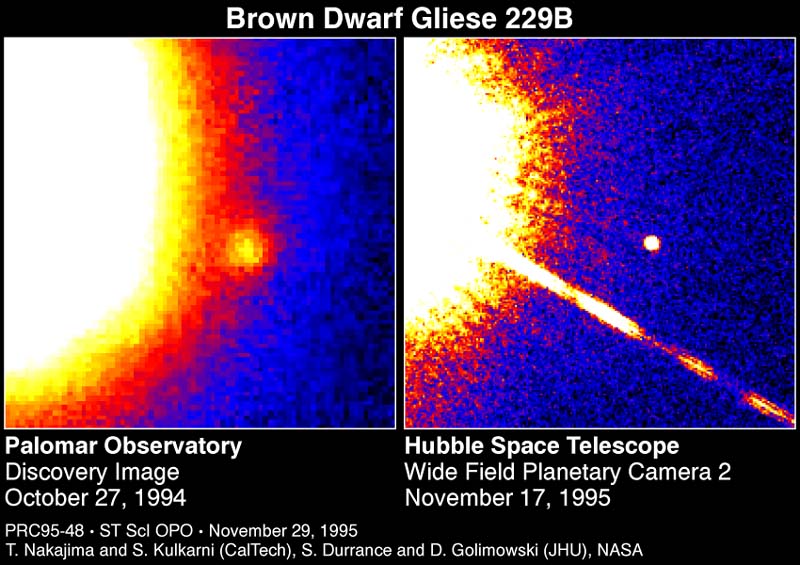}\hspace{0.5cm} 
\includegraphics[height=
5cm]{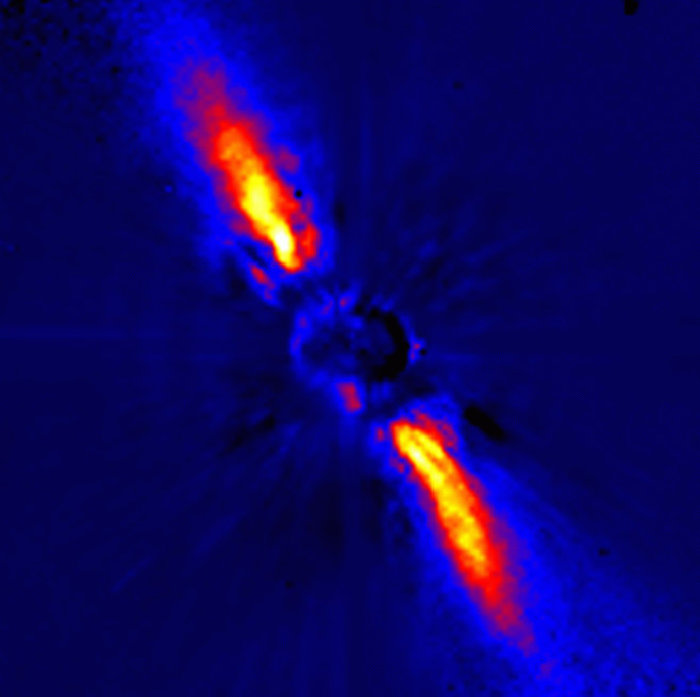} 
\caption{\textit{Left,} Discovery image of the cool brown dwarf companion Gliese 299\,B with the Johns Hopkins University AO Coronagraph at Palomar with the \textit{HST}/WFPC2 follow-up observation\cite{nakajima1995}. \textit{Right,} ESO3.6m/ADONIS coronographic observation of the edged-on debris disk around $\beta$ Pictoris\cite{mouillet1997}.}
\label{fig:precursors}
\end{center}
\end{figure}

%Heritage
Today's heritage in direct imaging (DI) of exoplanets is intimately connected to the pioneer work in the late 80's and early 90's for the development of Adaptive Optics (AO) system, infrared (IR) detectors, and coronographic techniques for the instrumentation of ground-based telescopes. The COME-ON AO prototype\cite{kern1989,rousset1990} (that will later become the ESO3.6m/ADONIS instrument\cite{beuzit1993}), the Johns Hopkins University AO Coronagraph\cite{golimowski1992}, or the CFHT CIRCUS coronographic camera at CFHT\cite{beuzit1991}, were precursor instruments that soon motivated the research and development of more sophisticated AO high-contrast imagers on 10m-class telescopes (PALAO-PHARO at Palomar, CIAO at Subaru, NIRC2 at Keck, NaCo at VLT, and NIRI and NICI at Gemini), confirming that ground-based instrumentation had demonstrated performances that could compete with space and \textit{HST}. Already at that time, following the discovery of the first brown dwarf GD\,165\,B\cite{zuckerman1992}, the power of combining high-angular resolution and high-contrast techniques was well envisioned for the discovery and characterization of substellar companions, including exoplanets, and protoplanetary and debris disks \cite{nakajima1994}. The emblematic discoveries and images of Gl\,229\,B\cite{nakajima1995} and $\beta$ Pictoris\cite{mouillet1997} shown in Fig.~\ref{fig:precursors} simply supported it, and motivated it even more. Later-on, with the dawn of exoplanet discoveries in radial velocity in 1995, DI started to routinely exploit 10m-class telescopes in the early 2000s to slowly joined the small family of planet hunting techniques known nowadays with radial velocity, transit, $\mu$-lensing and astrometry. Nowadays, DI currently brings a unique opportunity  to explore the
outer part of exoplanetary systems at more than 5-10~au to complete our view of planetary
architectures, and to explore the properties of relatively cool giant planets. The exoplanet's photons can indeed be spatially resolved and
dispersed to probe the atmospheric properties of exoplanets (and brown dwarf companions). As today's imaged exoplanets
are young (because they are hotter, brighter, thus easier to detect
than their older counterparts), their atmospheres show low-gravity
features, as well as the presence of clouds, and non-equilibrium
chemistry processes. These physical conditions are very different and
complementary to the ones observed in the atmospheres of field brown dwarfs or irradiated
inflated Hot Jupiters (studied in spectroscopy or with transit techniques like transmission
and secondary-eclipse, respectively). Finally, DI enables to directly probe the
presence of planets in their birth environment. Planet characteristics
and disk spatial structures can then be linked to study the planet --
disk interactions and the planetary system's formation, evolution, and stability, which is a fundamental and inevitable path to understand the formation of smaller telluric
planets with suitable conditions to host life.

%. Initial, developeed to chase also the BD vastly detected and characterized, performances enable to reconnect the with complementary 
%  techniques RV. These techniques are now , overlapping, but were for long separate communities.

%%%%%%%%%%%%%%%%%%%%%%%%%%%%%%%%%%%%%%%%%%%%%%%%%%%%%%%%%%%%%%%%%%%%%%%%%%%%%
\section{Telescope, instrument, techniques and observing strategies}
\label{sec:technics}  % \label{} allows reference to this section

The success of DI relies on a sophisticated instrumentation designed
to detect faint planetary signals angularly close to a bright star. A
Jupiter-like planet (orbiting at 5~au) around a typical young, nearby
star at 50~pc would lie at an angular separation of 100~mas setting the
order of angular separation we aim at. The typical planet--star
contrast are about $10^{-6}$ for a young Jupiter (today's
performances) and $10^{-8}$ for a mature Jupiter observed in emitted
light. It goes down $10^{-9}$ for a super-Earth in reflected-Light. From the recent discoveries of young,
massive self-luminous giant planets like $\beta$\,Pic\,b ($H$-band
contrast of $10^{-4}$ at 200-400~mas) with the first generation of
planet imagers at Palomar, Subaru, Keck, VLT, and Gemini to the first images of
an Exo-Earth with maybe an Extremely-Large Telescope (ELT), several
orders of magnitudes in contrast and separation must be
covered. Therefore, inovant technological developments are required to
meet the ultimate goal of imaging Exo-Earths (including the
construction of extremely large telescope and  the ability of achieving high-quality wavefront
control combined with ultimate coronographic and differential techniques). In that
perspective, the second generation of planet imagers, like the Spectro-Polarimetric
High-Contrast Exoplanet Resarch instrument\cite{beuzit2008} (SPHERE), the Gemini Planet Imager\cite{mac2008}
(GPI), and the Subaru Coronagraphic Extreme Adaptive Optics (SCExAO) instrument, aim at pushing high-contrast performances down to
$10^{-6}-10^{-7}$ at typical separation of 200~mas. They validate a dedicated instrumentation 
based on a 3-stages implementation design with: i/ high
angular resolution access, ii/ stellar light attenuation using
coronography, iii/ speckle subtraction using differential imaging
techniques. A fourth step can be added with the recent development of
powerful post-processing tools (iv/) to optimize the stellar signal
suppression. These instruments will undoubtedly serve as references and demonstrators
for future telluric planet imagers of the ELTs, including the proposed PFI\cite{macintosh2006} instrument for the TMT or EPICS\cite{kasper2008} for the European ELT .

\begin{figure}[t]
\begin{center}
\includegraphics[height=6cm]{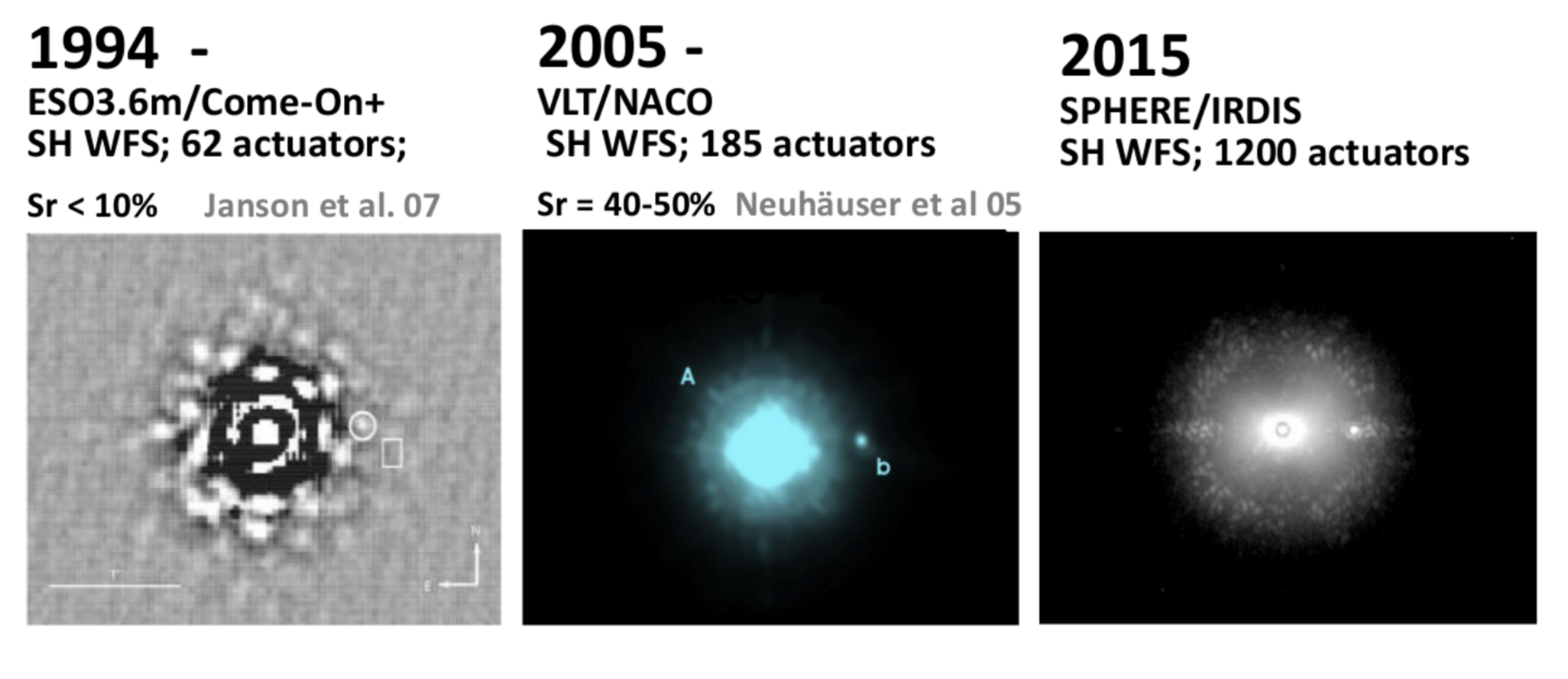} 
\caption{Observation fo the young GQ Lupi star (K7V; $V=11.4$; $K=7.1$) hosting a brown
  dwarf companion located at 730~mas with three generations of
  ESO AO instruments: ESO3.6m/Come-On+, NaCo and SPHERE at VLT.}
\label{fig:gqlupi}
\end{center}
\end{figure}

From the ground, the atmosphere turbulence affects the light
propagation and prevents large telescopes from reaching a spatial
resolution at the diffraction limit. Current Extreme-Adaptive
Optics (XAO) systems enable to compensate for atmospheric, but also
telescope and common path defects to routinely reach nowadays Strehl\footnote{The
  Strehl ratio can be defined as the ratio between the peak of the AO
  corrected PSF and the one of the ideal diffraction-limited Airy
  pattern.} correction of 90\% in H-band on bright ($V\le10$~mag)
targets. The evolution of the AO-systems performances over the past
decades is remarkable as illustrated in Fig.~\ref{fig:gqlupi}. Current XAO-instruments like SPHERE, GPI, and SCExAO rely on high-order deformable mirror, fast-temporal sampling frequency
at more than 1.0~kHz, spatial filtering of the wavefront before
sensing, tuned calibrations of the instrumental defects and enhanced
stability to limit low-order wavefront errors and alignment drift. The
main specifications are driven by the need to get i/ optimal spatial
sampling of the wavefront given the turbulence coherence length
(sampling better than $(D/r_0)^2$; $r_0\sim20$~cm at Paranal) and ii/
optimal temporal sampling to beat the turbulence speed (coherence time
$\tau_0$; 2-5\,ms at Paranal) to minimize the errors of wavefront
reconstruction and command. The correction stability in terms of
Strehl-correction, control of low-order aberrations and pointing
stability are essential to avoid any leakage or PSF deformation during
the observation that would significantly degrade the planet detection performances.

With a stable and diffraction-limited Point-Spread Function (PSF), the motivation for
coronography is relatively intuitive as it consists in blocking the
light from the central star to search for fainter objects in the close
stellar environment. Simple occulting masks or classical Lyot
coronographs (CLC) have been massively used in the past years on high-contrast imagers (see
Table~\ref{tab:planetimagers}) to
essentially: i/ reduce by a typical factor of $\sim100$ the intensity
of the central star diffracted-limited core and avoid any saturation
effects, ii/ reduce the intensity of the PSF wings without canceling the off-axis planetary signal, iii/ improve the observing efficiency by reducing overheads,
and iv/ reduce the total read-out noise. Classical and more recent apodized
version of the Lyot coronagraph are typically limited to
3--4$\lambda/D$ inner working angle (IWA)\footnote{The IWA is
  universally defined as the 50\% off-axis throughput point of a
  coronagraphic system}, mostly because they rely on amplitude
manipulation to attenuate starlight diffraction, covering a large area
of the PSF at the coronagraph plane. With the improvement of
instrumental stability achieved with the new generation fo planet
imagers, new concepts have emerged, particularly to access smaller
inner working angles down to $\lambda/D$ (see apodized vortex and
phase-induced amplitude coronographs). Pushing the instrumental
limitation to access smaller IWA down to $\lambda/D$ or $2\lambda/D$
(40 to 80~mas for the VLT at $H$-band) can make a significant
scientific breakthrough as the bulk of the giant planet population
around young, nearby stars is predicted to be located beyond the
ice-line at $\sim3$~au (i.e.  100~mas for a star at 30~pc). Various
concepts have been proposed and can be grouped in three main families:
i/ phase masks (four-quadran phase mask and vortex coronographs), ii/
phase/amplitude pupil apodization (phase-induced amplitude apodization
and apodizing phase plates) + focal plane masks (phase and/or
amplitude), and iii/ interferometers. Common to all these concepts is some
sort of phase manipulation, either in the focal plane or in the pupil
plane, or both\cite{mawet2012}. There is also a need to find a trade-off to mitigate
effects of chromatacity, throughput, sensitivity to low-order
aberrations considering ultimate contrast performances of each
concept. 
%Please refer to the review by  for further
%comparison of the various concepts of coronographs including classical
%and apodized Lyot coronographs and the phase-masks coronographs like
%the four-quadran-phase mask, currently implemented into SPHERE.

\begin{table}[t]
%\small
\begin{center}
\label{tab:planetimagers}
\caption{List of first and second planet imagers on 10-m class telescopes. FQPM/8QPM: four-quadrant phase mask, 8 octant phase mask. OVC: optical vortex corona-
graph. PIAAC: phase induced amplitude apodization. APP: apodizing phase plate.}
\begin{tabular}{llllll}     % 10 columns 
\noalign{\smallskip}\noalign{\smallskip}\hline\noalign{\smallskip}
Telescope & AO/Instruments        & $1^{st}$ Light & $\lambda$    &   Coronographs  & Diff. techniques \\
                        & & & ($\mu$m)        &                 &               \\
\hline\noalign{\smallskip}
Palomar  & PALAO/PHARO     & 2000     & $1.1-2.5$        & CLC/4QPM/OVC     & ADI \\
Subaru   & CIAO            & 2000     & $0.9-5.0$        & CLC                & ADI \\
Keck     & NIRC2-OSIRIS-NIRSPEC      & 2001     & $0.9-5.0$        & CLC/OVC          & ADI              \\   
VLT      & NACO            & 2002     & $1.0-5.0$        & CLC/FQPM/APP/OVC   & SDI/PDI/ADI      \\ 
Gemini-N & ALTAIR/NIRI     &  2003    & $1.1-2.5$        & CLC                & ADI             \\   
VLT      & MACAO/SINFONI         & 2004     & $1.0-2.5$        &                    & IFS      \\ 
Gemini-S & NICI            &  2007    & $1.1-5.0$        & CLC                & SDI/ADI          \\                   
Subaru   & AO188/HiCIAO    &  2008    & $1.1-2.5$        & CLC/PIAA/8QPM      & SDI/PDI/ADI      \\ 
Palomar  & PALAO/PHARO-P1640 & 2009     & $1.1-1.7$        & APLC/OVC          & IFS/ADI \\
LBT      & FLAO/LMIRCAM    & 2012     & $1.0-5.0$        & CLC/OVC            & ADI      \\
Magellan & MagAO/VisAO-CLIO& 2012     & $0.6-5.0$        & CLC                & ADI      \\  
Gemini-S & GPI             & 2013     & $1.1-2.3$        & CLC/ALC            & IFS/ADI/PDI      \\ 
VLT      & SPHERE          & 2014     & $0.5-2.3$        & CLC/ALC/4QPM       & IFS/ADI/PDI  \\ 
Subaru   & SCExAO/HiCIAO-CHARIS   & 2016    & $0.5-2.2$         & PIAA               & IFS/ADI/PDI      \\ 
\noalign{\smallskip}\hline                  
\end{tabular}
\end{center}
\normalsize
\end{table}

\begin{figure}[p]
\begin{center}
\includegraphics[width=12cm]{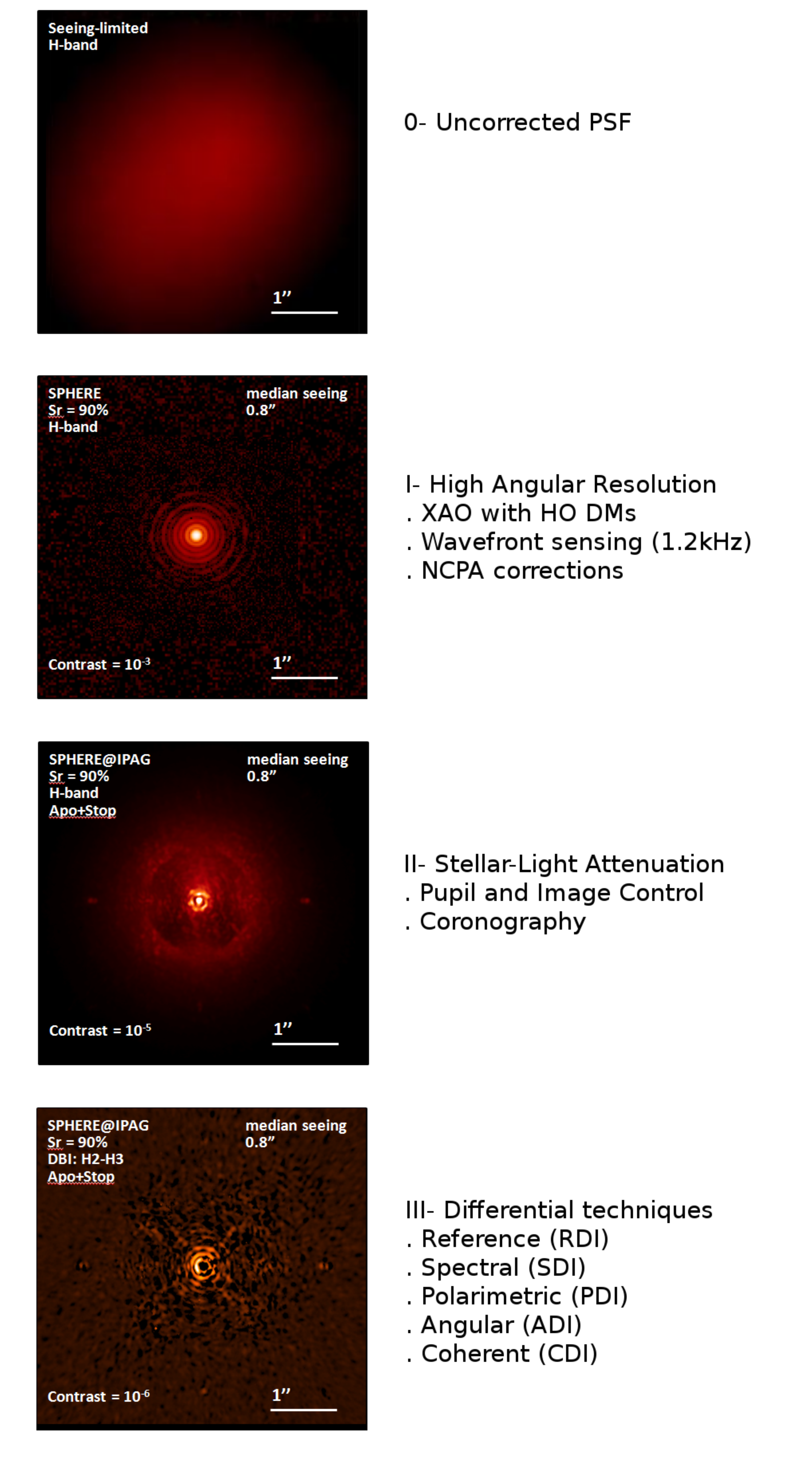} 
\caption{High angular resolution and high contrast implementation
  steps for AO imaging of exoplanets. Post-processing comes as a fourth step in the
  process for optimized PSF subtraction. NCPA refers to non-common path aberrations.}
\label{fig:challenge}
\end{center}

\end{figure}

As shown in Fig.~\ref{fig:challenge}, boiling atmospheric speckles and
instrumental quasi-statics speckles still remain an important source
of limitation in the XAO coronographic PSF. Differential imaging
techniques are nowadays a must-have for any high-contrast instruments
to address that issue. I will detail below the techniques commonly
used in the field of DI of exoplanets without considering Polarimetric
Differential Imaging (PDI) or a promising approach like Coherent
Differential imaging \cite{baudoz2012} (CDI) barely exploited for exoplanets for the moment. 

A classical approach widely used since the 90's for disk science with
\textit{HST} is to register an additional single star for reference
differential imaging (RDI). This technique has been successfully
applied to the case of $\beta$ Pictoris system and enabled the
detection of the $\beta$ Pic\,b planet with NaCo at VLT \cite{lagrange2009}. From the ground and with
AO-instruments, the reference must be well chosen to: i/ match
magnitude and colors of the science star to preserve the same AO
setting and the similar NIR flux, and ii/ match the parallactic angle
variation to keep a similar pupil configuration (for observations with
alt-az telescopes). With an increased stability of the
telescope+instrument and a duty cycle of a few minutes, high degree of
correlation may in addition persist between the science star and the
reference to significantly gain in contrast. This promising technique
of star hopping still needs to be validated in terms of performances
with the second generation of planet imagers if (pointing, AO setting)
overheads can be reduced.

An alternative differential technique makes use of the planet spectral
properties.  Giant planet atmospheres are composed of chemical
elements (water, methane, amonia, carbon monoxide and/or dioxide), not
present in the stellar atmospheres. Signatures of planetary accretion in emitting lines like H$\alpha$ can also be used\cite{close2014}. These differential spectral
signatures between the planet and the star can be optimally used to
suppress part of the stellar light and to reveal broad molecular
absorptions or accretion lines in the planet spectrum. The spectral differential imaging
technique (SDI) \cite{racine1999,marois2000} relies on the simultaneous observations of the
star+planet at well defined wavelengths (on/off defined molecular
absorptions quoted above). Although it efficiently tackles the
subtraction of residual atmospheric speckles, SDI remains: i/
sensitive to differential aberrations between the different imaging
paths and ii/ optimized for the detection of cool or accreting, and spectrally contrasted
planets. An extension of this approach makes use of integral
field spectrographs offering a broader spectral range. Optimized
subtraction of the PSF simultaneously observed at different
wavelengths considering the complete star/planet NIR spectral
properties enables a significant gain in terms of detection
performances. Two SPHERE instruments, the IRDIS Dual-Band image and
the IFS Integral Field Spectrograph have been designed to exploit that
concept, as the GPI and SCExAO/CHARIS planet imagers which have for unique instrument an integral field spectrograph.

Finally, a last and complementary technique exploits the field and
pupil rotation of alt-az Telescope. This is the so-called angular
differential imaging technique\cite{marois2006} (ADI). It enables
to discriminate quasi-statics speckles (over the duration of the
observations) and fixed instrumental aberrations (frozen when the
pupil is stabilized) from astrophysical sources like exoplanets
rotating with the field. The efficiency of that technique highly
depends on the parallactic angle variation rate (therefore the target
coordinates and observing time) and the companion separation given the
resolving power of the telescope. ADI can be optimally applied after
SDI to remove differential aberrations between the SDI paths. The
SPHERE and GPI imagers actually combine both techniques to boost their
detection performances, however care must be taken as they affect the
planetary (astrometric and spectro-photometric) signal.
  
With these various stages of instrumental complexity (XAO, low order
aberration control, coronography and differential techniques),
important progress in the past years has been made to develop
innovative algorithms to optimally calibrate the PSF, spatially,
temporally and spectrally in order to suppress the stellar
contribution and recover faint planetary signals as close as possible
to the star. These algorithms make heavy use of the known modulation
of the signal imprinted by the observing strategy. They
deploy various degrees of complexity for example in the context of ADI
with: i/ classical ADI (cADI) using the median of the ADI cube as a
reference PSF, ii/ smart ADI (sADI) with a reference PSF selection
criteria for each science image considering frames that have rotated
by more than $\alpha\times$FWHM ($\alpha$ being a resolution criteria
used as input), iii/ radial ADI (rADI) following
the reference PSF selection criteria of sADI, but repeated for various
separations, iv/ locally optimized combination of images\cite{lafreniere2007} (LOCI) with the definition of optimization zone
(larger than the subtraction zone) where the coefficients for the
linear optimization are computed by minimizing the sum of the squared
residuals of the subtraction using various set of parameters (criteria
of separation, size of optimization zone, radial width and ratio of
the radial-to-azimuthal widths of subtraction zones) and v/ principal
component analysis\cite{soummer2012,amara2012} (PCA) that uses an
orthogonal basis of eigenimages, on which the science target is
projected to create the PSF reference. Spectrally, whereas simple
spatial and flux rescaling are used in SDI, more elaborated
approach like spectral deconvolution \cite{sparks2002}, spectral
LOCI and PCA \cite{pueyo2012,mesa2015}, forward modeling for specific characterization\cite{greenbaum2018} or more recently combined high or even medium-resolution dispersed spectroscopy with molecular mapping  \cite{hoeij2018} can also be applied to
optimize the PSF subtraction and minimize the planetary signal
cancellation using Integral Field Spectrograph to exploit the spectral diversity.

Whereas the flux subtraction in ADI processing is purely geometric
and related to the planet position and parallactic angle variation
during the observation, the situation is much more complex for SDI.
The SDI signature of any
point-source depends on the separation, luminosity, but also on the
spectral properties of the object. Self-subtraction cannot be reduced to a simple geometric effect. As a
consequence, the detection performances of SDI observations (and more generally high and medium-resolution dispersed spectroscopy coupled to AO) cannot be
expressed as a contrast in magnitude relative the central star without an a-priori
knowledge of the spectral properties of the searched companions
\cite{rameau2015}. Consequently, although the
characterization of detected planets can be individually and locally
treated using injection of fake planets in the data in ADI, SDI or
ASDI (combination of ADI and SDI), the determination of accurate
detection limits requires further a priory assumptions on the spectral
properties of the hunted planets.

%%%%%%%%%%%%%%%%%%%%%%%%%%%%%%%%%%%%%%%%%%%%%%%%%%%%%%%%%%%%%%%%%%%%%%%%%%%%%
\section{Targeting young, nearby stars}
\label{sec:targets}  % \label{} allows reference to this section

Young ($\le500$~Myr), nearby ($\le 100$~pc) stars have been prime targets
for DI of exoplanets for more than a decade now. The identification of
these comoving groups of young stars started from an anomaly observed
three decades ago. The TW\,Hya T\,Tauri star was discovered isolated
from any dark cloud nor birth place regions \cite{rucinski1983}. A
few years later, IRAS excesses combined with H$\alpha$ and Lithium depletion measurements enabled the
identification a handful of young stars in the vicinity of TW\,Hya. Their
origin, runaway stars from some cloud or formed in situ from a
low‐mass cloud, their age and their distance remained unclear for
years. X-ray data finally confirmed that they were members of the
first identified young, nearby association, i.e. the nearest known
region of recent star formation, the so-called TW Hydrae association\cite{kastner1997}
(TWA). Over the next decade, the number of TWA
members drastically increased to more than 30 members, with age and
mean distance estimates converging towards 10~Myr and 50~pc,
respectively. Immediately after the confirmation of the existence of
TWA, several research groups became specifically interested in
identifying new young, nearby associations in the vicinity of the Sun
\cite{zuckerman2004,torres2008}.

In most current studies, the identification relies on two main pillars:
\begin{itemize}
\item Galactic kinematics studies (space velocity
  analyses). Techniques have become increasingly sophisticated using
  for instance Bayesian analysis to derive membership probability
  considering combined photometric and kinematics catalogs,
\item Youth diagnostics determination. Various diagnostics can be used
  depending on the stellar age and spectral type like color-magnitude
  diagram, rotation, Lithium depletion, H$\alpha$ emission, Ca H\&K,
  Coronal X-rays and chromospheric UV activity, or the presence of IR
  excess to inter-compare and refine stellar ages.
\end{itemize}

Since the discovery of TW Hydrae, more than 500 young, nearby ($< 100$\,pc) stars
were identified, mostly F, G, K and M dwarfs. Recent systematic studies
are now pushing the horizon down to very low-masses with the discovery
of a large population of M-, L- and even T-type members
\cite{gagne2015}. These young, nearby stars are gathered in several
groups (TW Hydrae, $\beta$ Pictoris, Tucana-Horologium, $\eta$ Cha, AB
Dor, Columba, Carinae...), sharing common kinematics, photometric and
spectroscopic properties. They are also more recently completed by stars with intermediate-age or located at larger distance like in the so-called Sco-Cen complex. Their main
properties are reported in Table~\ref{tab:yns}. With typical contrast of
$10-15$ magnitudes for separations beyond $1.0-2.0~\!''$ ($50-100$ au
for a star at 50~pc) achieved with the first generation of planet
imagers, planetary mass companions down to $1-2$ Jupiter masses became
potentially detectable. Therefore, without surprise, a significant
amount of telescope time in the early-2000's was dedicated to deep
imaging surveys of young, nearby stars to search for exoplanets, brown
companions and disks leading to the first discoveries of planetary
mass companions in DI.

%%%%%%%%%%%%%%%%%%%%%%%%%%%%%%%%%%%%%%%%%%%%%%%%%%%%%%%%%%%%%%%%%%%%%%%%%%%%%
\section{Discoveries and surveys}
\label{sec:discoveries}  % \label{} allows reference to this section

\begin{figure}[t]
\begin{center}
\includegraphics[height=7cm]{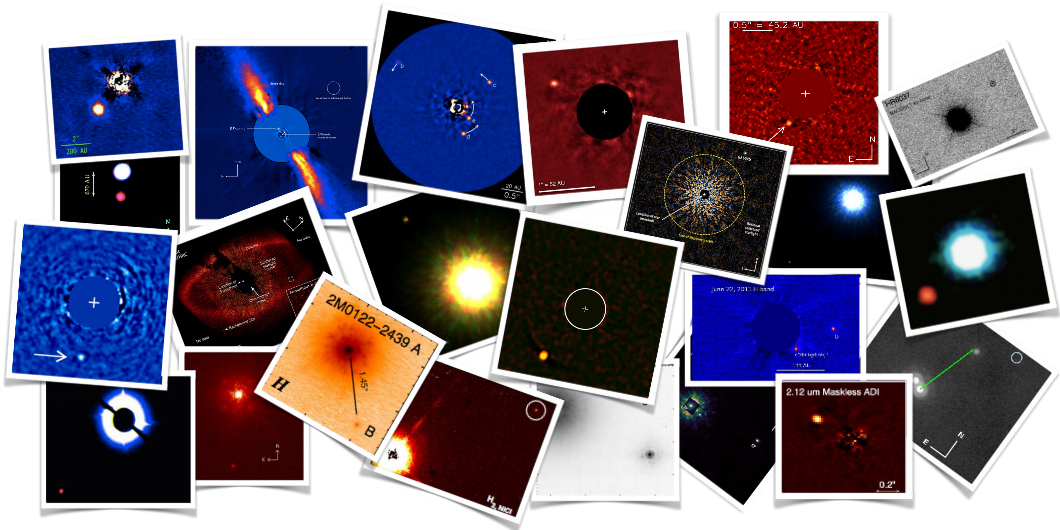} 
\caption{Non-exhaustive family portrait of exoplanets and/or planetary mass companions discovered in DI including the discoveries of (from right to left): DH\,Tau\,b, UScoCTIO\,108\,b, 51 Eri\,b, AB Pic\,b, $\beta$ Pic b, Fomalhaut\,b, 2M0122-2439\,b, HR\,8799\,bcde, RXJ1609 b, HD\,106906\,b, $\kappa$ And\,b , HIP65426\,b, CD-35 2722\,b, GJ\,504\,b, HD\,95086\,b, GU Psc\,b, HIP\,77900\,b, HR\,6037\,b, 2M1207\,b and 2M0143AB\,b.}

\label{fig:family}
\end{center}
\end{figure}
In the course of a DI survey, the discovery of a serious planetary
candidate is usually indicated by a combination of small separation,
high contrast and cool atmosphere features. Very red near-IR colors or specific
spectral features such as a peaked H-band spectrum are indicative of
young L dwarfs atmospheres. Methane absorption detected in SDI
indicates a probable young and cool T dwarf. These diagnostics enable
the observers to rapidly estimate the predicted mass,
effective temperature and physical separation of the candidate using
evolutionary model predictions. It highlights one key limitation of DI
which is the dependency on luminosity-mass predictions from
uncalibrated formation and evolutionary models detailed here after.

\begin{table}[t]
\begin{center}
\label{tab:planets}
\caption{Non-exhaustive, but illustrative list of substellar/planetary-mass companions and/or exoplanets discovered by DI around young, nearby stars in the last two decades.}
\begin{tabular}{lllllllllll}     % 10 columns
\noalign{\smallskip}\hline  \noalign{\smallskip}
\multicolumn{5}{c}{Primary} & \multicolumn{5}{c}{Companion} & \\
%\cmidrule(l){1-5} \cmidrule(l){6-10} 
Name                & Asso.    & Age      &  d        & SpT      & $\rho$  & $\Delta$ & $\Delta H$ & Mass$^a$        & SpT      & Date   \\
                    &          & (Myr)    & (pc)      &          & (as)    & (AU)     & (mag)      & M$_{\rm{jup}}$ &         &       \\         
\noalign{\smallskip}\hline  \noalign{\smallskip}
Gl\,229 \,B             & Field      &  $\sim500$     &     5.8    & M1V        &   7.8   & 45      &   10.0    
&     35      &  T6.5   &  1995  \\
...            &       &       &        &         &      &       &     &           &     &    \\
TWA\,5 \,B             & TWA      & 10       & 50        & M3       & 2.0     & 100      & 4.9       & 20           & M9     &  1999        \\
HR\,7329\,B            & $\beta$ Pic& 20     & 48        & A0       & 4.0     & 200      & 6.9       & 20           & M8     & 2000      \\
GSC\,08047\,B          & Tuc-Hor  & 40       & 85        & K1       & 3.2     & 250      & 6.9    & 25           & M8     & 2003      \\
2M1207\,b              & TWA      & 10       & 53        & M8       & 0.78    & 41       & 5.7    & 3            & L3     & 2004      \\
AB\,Pic\,b             & Tuc-Hor  & 40       & 45        & K1       & 5.5     & 248      & 7.6    & 12           & L1     & 2005      \\  
GQ\,Lup\,b             & Lupus    & 1        & 140       & K1       & 0.7     & 100      & 6.0       & 12           & M8     & 2005 \\
DH\,Tau\,b             & Tau      & 2        & 140       & M7       & 2.3     & 330      & 6.2     & 15           & M8      & 2005 \\
CHXR\,73\,b            & Cha      & 2        & 160       & M        & 1.3     & 210      & 5.2       & 12           & M8     & 2006   \\ 
CT\,Cha\,b             & Cha      & 2        & 165       & M        & 2.7     & 440         & 5.5       & 15           & M8      & 2008 \\
RXJ1609\,b             & Upp Sco  & 5        & 150       & K7       & 2.2     & 330      & 7.8        & 8            & L4     & 2008 \\
HR\,8799 (b)        & Col      & 30       & 39.4      &  A5      & 1.72    & 68       & 12.5       & 7            & L/T    & 2008 \\ 
  \hspace{1.45cm}(c) &   -      &  -       & -         & -        & 0.94    & 37       & 11.6       & 10           & L/T    & 2008\\ 
  \hspace{1.45cm}(d) &   -      &  -       & -         & -        & 0.66    & 26       & 11.6       & 10           & L7     & 2008\\ 
  \hspace{1.45cm}(e) &   -      &  -       & -         & -        & 0.39    & 15       & 11.5       & 10           & L7     & 2010 \\ 
$\beta$ Pic\,b         & $\beta$ Pic & 23     & 19.3     & A6       &0.4      & 9        & 10.0       & 8            & L1     & 2008 \\ 
Fomalhaut\,b           & Field       & 200-400& 7.7      & A4       & 15.5        & 115  & -          & $\le3$       & ?      & 2008 \\ 
$\kappa$ And\,b        & Col?        & 10-150 & 51.6     & B9       &  1.1       & 55         & 10.6           & 13             &  L5      & 2013 \\ 
HD\,95086\,b           & LCC         & 17     & 90.3     & A8       & 0.6     & 61       & 13.1       & 5            & L7    & 2013 \\ 
GJ\,504\,b             & Field       & 120    & 17.5     & G0       & 2.5        &  44        & 16.3           & 4             & T      & 2013 \\ 
GU Psc\,b           & AB Dor      & 140    & 48       & M3       & 42      & 2000     & 8.1        & 11           & T3.5   & 2014 \\ 
51 Eri\,b              & $\beta$ Pic & 23     & 29.4     & F0       & 0.55    & 13       & 14.4       & 2            & T3     & 2015 \\ 
HIP\,65426\,b     & LCC & 17     & 83.9     & A2       & 0.83    & 92       & 11       & 9            & L5     & 2017 \\ 
PDS\,70 \,b    & UCL & 3     & 113.4     & K7       & 0.19    & 9.3       & 29       & 5            & mid-L     & 2018 \\ 
\noalign{\smallskip}\hline                  \noalign{\smallskip}
\end{tabular}
\end{center}
\normalsize
\end{table}

In the case of the 2M1207\,b planetary mass companion, the very red
near-IR colors ($K-L\,\!'=1.65\pm0.20$~mag) were indicative of a very
dusty atmosphere. DUSTY models were predicting a mass of 3--5~M$_{\rm{Jup}}$
\cite{chabrier2000}. Firmed confirmation in DI then came with
follow-up test at different epochs to verify that the candidate was
actually co-moving with the central star. Orbital motion can be resolved to unambiguously
confirm that both objects are physically bound. Depending on the
parallactic and proper motion of the central source as well as the
astrometric precision of the planet imager, up to one year interval
between both epochs could be necessary for final confirmation as
illustrated in the case of 2M1207 ($\mu_{\alpha}=-78\pm11$~mas/yr,
$\mu_{\delta}=-24\pm9$~mas/yr; $d=53.2$~pc; typical NaCo astrometric
precision $\sim10$~mas). With GPI, SPHERE, SCExAO reaching an astrometric precision
down to 1-2~mas, a few weeks to a few months are now sufficient, which shows
 the gain of enhanced instrumental performances. After the discovery of the first brown dwarf companion Gl\,229\,B\cite{nakajima1994} in
1994 that benefited from the combined technological innovation of high contrast techniques, 
the ones of the first young brown dwarf companions (TWA5\,B\cite{lowrance1999} and HR\,7329\,B\cite{lowrance2000}) in young,
nearby associations that exploited the \textit{HST/NICMOS} stability, the first
generation of planet imagers on 10m-class telescopes at Palomar, Subaru, Keck, VLT, Gemini, and later-on LBT and Magellan enabled systematics survey of large
sample of young, nearby stars. They rapidly led to the discovery of
the first planetary mass companions in the early 2000's. The first
companions were detected at large distances ($\ge100$~au) and/or with
small mass ratio with their primaries, indicating a probable formation
via gravo-turbulent fragmentation
\cite{hennebelle2011} or gravitational disk instability as already addressed early-on in the discovery papers \cite{chauvin2005b,lafreniere2008}.
The implementation of differential techniques, starting in 2005--2006,
enabled the breakthrough discoveries of closer and/or lighter
planetary mass companions like HR\,8799\,bcde\cite{marois2008,
  marois2010} (10, 10, 10 and
7~M$_{\rm{Jup}}$ at resp. 14, 24, 38 and 68~au), $\beta$\,Pictoris\,b\cite{lagrange2009} (8~M$_{\rm{Jup}}$ at 8~au), Fomalhaut~b\cite{kalas2008} ($<1$~M$_{\rm{Jup}}$ at 177~au; still debated), $\kappa$~And\,b\cite{carson2013} ($14^{+25}_{-2}$~M$_{\rm{Jup}}$ at 55~au), HD\,95086\,b\cite{rameau2013b} ($4-5$~M$_{\rm{Jup}}$ at 56~au), GJ\,504\,b\cite{kuzuhara2013}
($4^{+4.5}_{-1}$~M$_{\rm{Jup}}$ at 43.5~au),
51\,Eri\,b\cite{macintosh2015} ($2$~M$_{\rm{Jup}}$ at 13~au), HIP\,65426\,b\cite{chauvin2017} ($9$~M$_{\rm{Jup}}$ at 92~au), and more recently PDS\,70\,b ($9$~M$_{\rm{Jup}}$ at 29~au, Keppler et al. 2018, accepted). They
indicate that we are just initiating the characterization of the giant
planet population at wide orbits, between typically
$5-100$~au. Fig.\,\ref{fig:family} and Table~\ref{tab:planets} show and summarize the non-exhaustive, but illustrative family
portrait of young very low mass brown dwarf companions and/or planets
discovered in DI in the past two decades.

\begin{table}[t]
%\small
\begin{centering}
\label{tab:largesurveys}
\caption{Deep imaging surveys of young ($<100$~Myr) and
  intermediate-old to old ($0.1-5$~Gyr), nearby ($<100$~pc) stars
  dedicated to the search for planetary mass companions. We have
  indicated the telescope and the instrument, the imaging mode (I: standard imaging; Sat-I; saturated imaging; Cor-I:  coronagraphic imaging ; SDI:
  simultaneous differential imaging; ADI: angular differential
  imaging; ASDI: angular and spectral differential imaging), the
  filters, the field of view (FoV), the number of stars observed (\#),
  their spectral types (SpT) and ages (Age). }
\begin{tabular}{lllllllllll}     % 10 columns 
\hline\noalign{\smallskip}
Reference               & Telescope        & Instr.       &  Mode          & Filter       & FoV          & \#       & SpT      & Age       \\ 
                        &                  &              &                &              & (as)     &          &          & (Myr)     \\
\noalign{\smallskip}\hline\noalign{\smallskip}
Nakajima+94     & Palomar          & AOC       & Cor-I          &    $I$     & $60$ & 24      & G-M      & Field  \\ 
...     &          &        &          &         &  &      &       &   \\ 
Chauvin+03     & ESO3.6m          & ADONIS       & Cor-I          & $H, K$        & $13$ & 29      & G--M      & $\lesssim50$  \\ 
Neuh\"auser+03  & NTT              & Sharp/Sofi        & Sat-I          & $K/H$          & $20$ & 33       & A--M     & $\lesssim50$  \\ 
Lowrance+05    & \textit{HST}              & NICMOS       & Cor-I          & $H$          & $19$ & 45       & A--M    & $10-600$  \\
Masciadri+05   & VLT              & NaCo         & Sat-I          & $H, K$        & $14$ & 28       & KM       & $\lesssim200$  \\ 
Biller+07      & VLT/MMT              & NaCo/ARIES         & SDI            & $H$          & $5$   & 45       & G--M      & $\lesssim300$  \\
Kasper+07      & VLT              & NaCo         & Sat-I          & $L'$         & $28$ & 22       & G--M      & $\lesssim50$  \\ 
Lafreni\`ere+07& Gemini-N         & NIRI         & Sat-ADI            & $H$          & $22$ & 85   &  F--K        & 10-5000   \\
Apai+08        & VLT              & NaCo         & SDI            & $H$          & $3$   & 8    &    FG & 12-500 \\
Chauvin+10     & VLT              & NaCo         & Cor-I          & $H, K$        & $28$ & 88       & B--M    & $\lesssim100$          \\  
Heinze+10ab    & MMT              & Clio         & Sat-ADI            & $L', M$       & $15.5$ & 54   & F--K      & 100-5000  \\                       
Janson+11      & Gemini-N         & NIRI         & Sat-ADI            & $H, K$        & $22$ & 15       & BA       & 20-700    \\  
Vigan+12       & Gemini-N/VLT         & NIRI         & Sat-ADI            & $H, K$        & $22/14$ & 42       & AF       & 10-400    \\ 
Delorme+12     & VLT              & NaCo         & Sat-ADI            & $L'$         & $28$ & 16       & M        & $\lesssim200$  \\  
Rameau+13c      & VLT              & NaCo         & Sat-ADI            & $L'$         & $28$ & 59       & AF       & $\lesssim200$  \\
Yamamoto+13       & Subaru           & HiCIAO       & Sat-ADI            & $H, K$          & $20$ & 20       & FG       & $125\pm8$  \\
Biller+13      & Gemini-S         & NICI         & Cor-ASDI       & $H$          & $18$ & 80       & B--M     &  $\lesssim200$    \\ 
Nielsen+13     & Gemini-S         & NICI         & Cor-ASDI       & $H$          & $18$ & 70       & BA       & 50-500    \\ 
Wahhaj+13      & Gemini-S         & NICI         & Cor-ASDI       & $H$          & $18$ & 57       & A--M    & $\sim100$ \\ 
Janson+13      & Subaru           & HiCIAO       & Sat-ADI            & $H$          & $20$ & 50       & A--M    & $\lesssim1000$ \\  
Brandt+14      & Subaru           & HiCIAO       & Sat-ADI            & $H$          & $20$ & 63       & A--M    & $\lesssim500$    \\ 
Chauvin+15      & VLT           & NaCo       & Sat-ADI            & $H$          & $14$ & 86       & F--K    & $\lesssim200$ \\  
Meshkat+15ab     &  VLT  & NaCo  & APP-ADI  &  $L'$   & 28 & 20  & AF & $\lesssim200$  \\
Bowler+15       & Keck/Subaru   & NIRC2/HiCIAO  & Cor-ADI  & $H$      & 10/20  &  78 & M & $\lesssim200$ \\
Galicher+16     &  Keck  & NIRC2  & Cor-ADI  &  $H,K$   & 10 & 229  & A-M & $\lesssim200$  \\
&  Gemini-N/S  & NIRI/NICI  &   &    & &   &  &   \\
Durkan+16     & \textit{Spitzer}  & IRAC  & I & $4.5\,\mu$m    & 312 &  73 &  A--M &  $\lesssim200$ \\
\noalign{\smallskip}\hline                  
\end{tabular}
%\begin{list}{}{}
%\item[\scriptsize{- ($^a$):}] \scriptsize{surveys dedicated to planets around debris disk stars.}
%\end{list}
\end{centering}
\normalsize
\end{table}

The dozens of discoveries obtained in DI so far has been possible
thanks to large surveys of young stars that mostly
identified background contaminants (see Table~\ref{tab:largesurveys}). Various motivations were
followed for the target selection of these surveys: i/ complete census
of given associations, ii/ selection of young, intermediate-mass stars, iii/ or very
low-mass stars, or iv/
application of figure of merit considering detection rate with toy
models of planet population. The near-infrared
wavelengths ($H$ and $K_s$) have been used intensively. They are a good
compromise between low-background noise, angular resolution and good
Strehl correction. However, thermal imaging has been very
competitive in terms of detection performances as the planet-star
contrast and the Strehl correction are more favorable in those
wavelengths despite an increased thermal background. For instance,
SPHERE is less performant than NaCo for the detection of giant planets
around young, nearby M dwarfs at typical separation larger than
500-1000~mas. The low-rate of planet detection in DI considering the large number of
stars observed has motivated the approach of new systematic and very
large scale surveys (as conducted in radial velocity for instance).
The Gemini GPIES survey and the VLT/SPHERE SHINE\cite{chauvin2017b} (The SpHere INfrared 
survey for Exoplanets) will observe both 600 stars over four to five years. 
With enhanced detection performances, the objective is to significantly
increase the number of imaged planets to characterize, but also to
provide better statistical constraints on the occurrence and the
characteristics of exoplanets at wide orbits ($\ge5$~au). This will
give us a more global picture of planetary systems architecture at all
orbits to improve our understanding of planetary formation and
evolution mechanisms. Early-results already confirm the gain in terms
of detection performances compared with the first generation of planet
imagers.

%\begin{figure}[t]
%\begin{center}
%\includegraphics[height=10cm]{output_5c.eps} 
%\caption{Exoplanets discoveries in May 2016 considering the different planet hunting
%  techniques. The exoplanet masses are reported as a function of their
%  distances to the star.}
%\label{fig:diagplanets}
%\end{center}
%\end{figure}

%%%%%%%%%%%%%%%%%%%%%%%%%%%%%%%%%%%%%%%%%%%%%%%%%%%%%%%%%%%%%%%%%%%%%%%%%%%%%
\section{Key astrophysical results}
\label{sec:results}  % \label{} allows reference to this section

%%%%%%%%%%%%%%%%%%%%%%%%%%%%%%%%%%%%%%%%%
\subsection{Mass \& accretion history}
\label{subsec:mass}

\begin{figure}[t]
\begin{center}
\includegraphics[height=7cm]{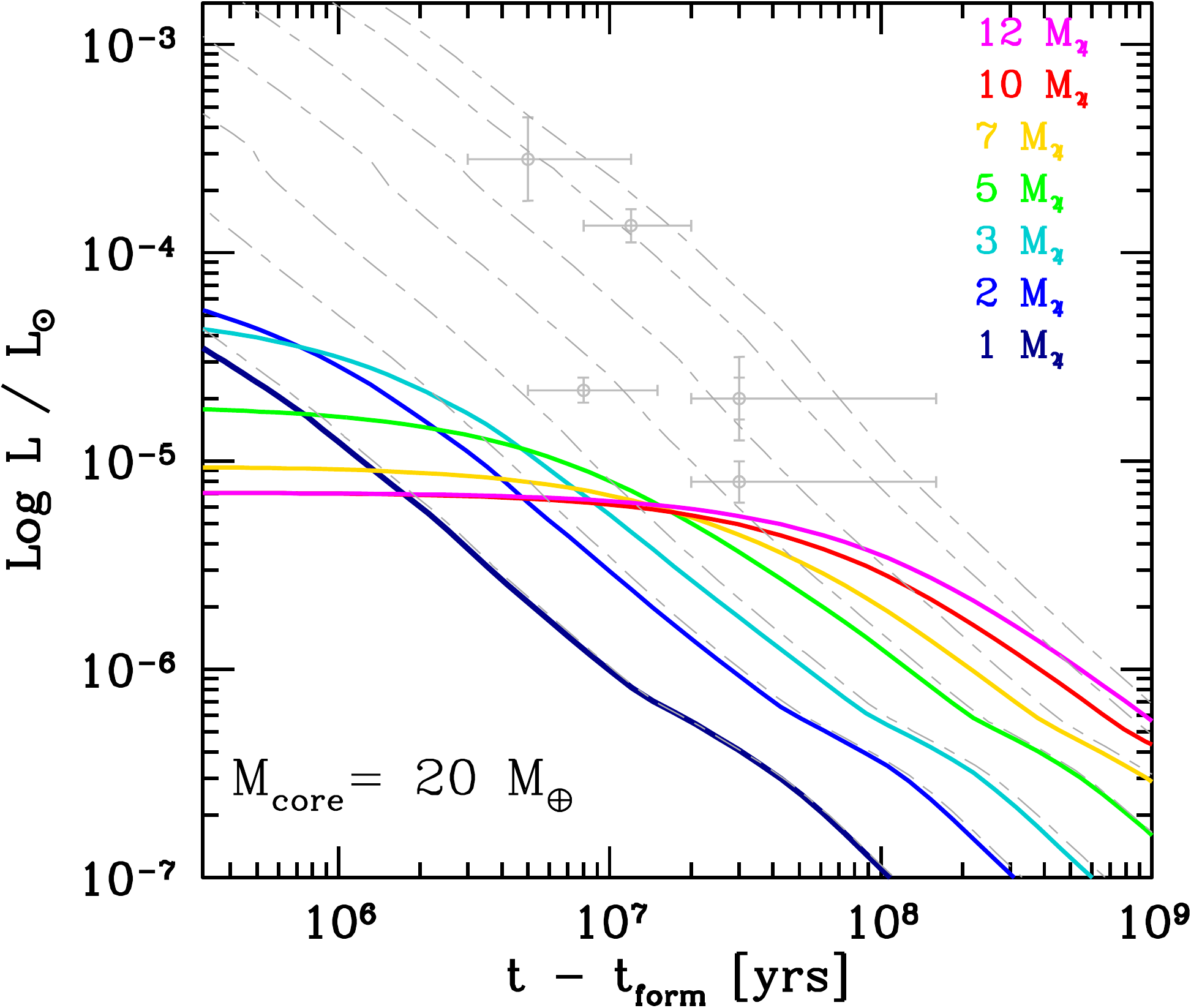} 

\caption{Luminosity as a function of time after formation for planets
  with total masses of 1, 2, 3, 5, 7, 10, and 12 M$_{\rm{Jup}}$ and core masses of
  20, 33, 49, and 127 M$_{\oplus}$ as indicated in the panels based on the BEX  models\cite{mordasini2013}. The colored solid
  lines assume cold-start cases. The dashed-dotted gray lines assume hot-start cases. The points with error bars are young giant planets (from
  top to bottom: RXJS\,1609\,b, $\beta$\,Pic\,b, 2M1207\,b, HR\,8799\,c, d, e and b).}

\label{fig:evol}
\end{center}
\end{figure}

As previously mentioned, DI gives access to the photometry, the luminosity and the spectral energy distribution of
exoplanets, but not directly to the mass. In this context, we have to
rely on evolutionary model predictions that are not well calibrated at
young ages. In addition to the system age uncertainty, the predictions
highly depend on the formation mechanisms and the gas-accretion phase
that will form the exoplanetary atmosphere. The way the accretion shock will
behave (sub or super-critical) on the surface of the young accreting
proto-planets during the phase of gas runaway accretion will drive the
planet initial entropy or internal energy, hence its initial physical
properties (luminosity, effective temperature, surface gravity and
radius) and their evolution with time. These different physical states
are described by the so-called hot-start (sub-critical shock),
cold-start (super-critical shock), and warm-start (intermediate
cases) models. They predict luminosities that are spread over several
order of magnitudes for young, massive giant planets. This is
illustrated by Fig.~\ref{fig:evol} based on the BEX models\cite{mordasini2013}
where predictions in terms of luminosity are reported for different
planet masses as a function of time and for the cold-start and hot-start cases. The higher is the mass,
the larger is the spread and the time duration for the models to
converge. Nowadays, most masses reported in the literature or in
table~\ref{tab:planets} for imaged planets are the ones predicted by
hot-start models and may under-estimate the planet masses.

\begin{figure}[t]
\begin{center}
\includegraphics[height=5.5cm]{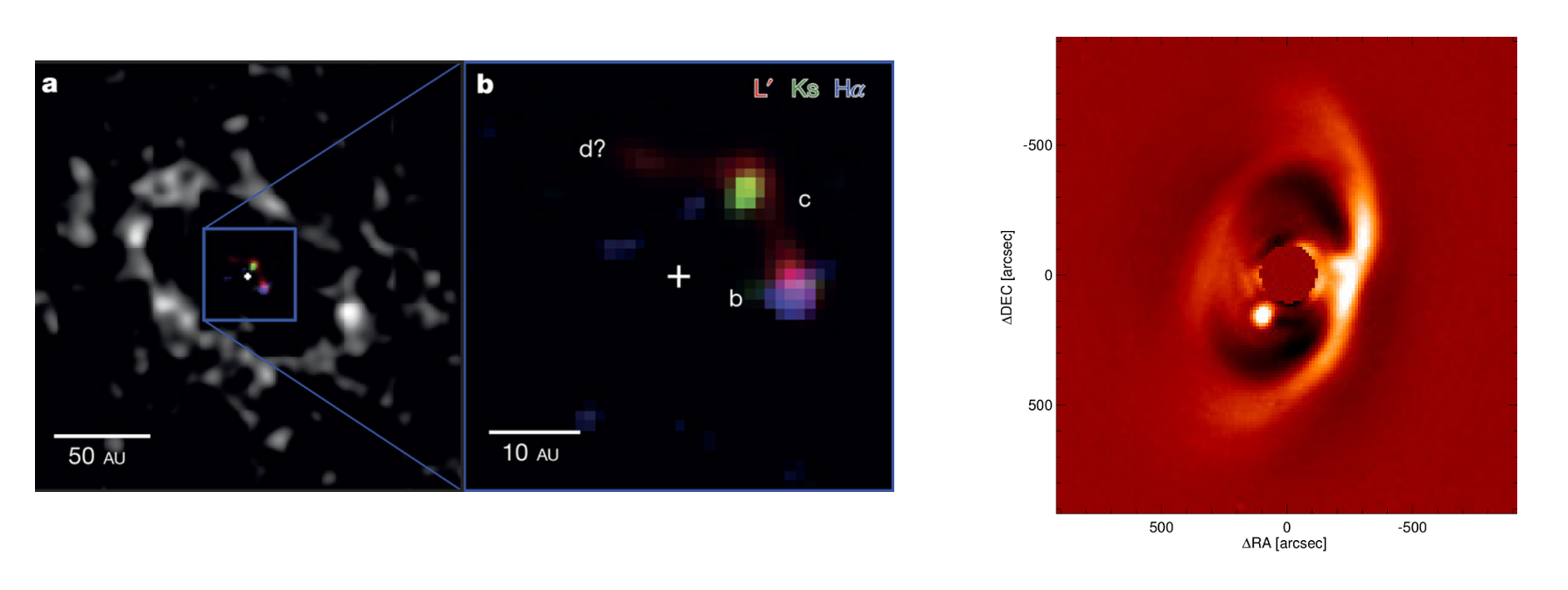} 

\caption{\textit{Left:} Composite $H_\alpha$, $Ks$, and $L~\!'$ image of accreting
  protoplanets in the LkCa 15 transition disk. The coloured
  image shows $H_\alpha$ (blue), $Ks$ (green), and $L~\!'$ (red)
  detections at the same scale as VLA millimetre observations
  (greyscale). An expanded view of the central part of the
  cleared region is also showing the composite image of LBT and Magellan
  observations, with b, c, and d proto-planets\cite{sallum2015}.
  \textit{Right:} VLT/SPHERE-IRDIS ADI image of the circumstellar environment 
  around the young, nearby star PDS\,70 
  (K7, 3\,Myr, 113\,pc, UCL region) revealing the transition disk and the 
  young exoplanet PDS70\,b in $H2H3$-band (Keppler et al. 2018, accepted).}

\label{fig:proto}
\end{center}
\end{figure}

The consequence is that much remain to be understood about the
formation of giant planet and of their atmospheres from the theoretical point of
view. The observations of proto-planets in their birth environment
and the direct measurement of mass accretion rate using $H\alpha$
differential imaging would add interesting constraints on the way gas
is accreted on young forming planets (see for instance the pioneer case of HD\,142527\,B
detected using MagAO\cite{close2014}). Several interesting
candidates in transition disks will maybe shed some light on this
\textit{terra incognita} in terms of mass and accretion rate regime
(see Fig.~\ref{fig:proto}). One exciting solution comes from the combination of DI and radial
velocity and/or astrometry to derive the dynamical mass
of imaged planets with their luminosity in order to explore the possible states of initial entropies
that are consistent with observations. Future systematic and combined characterization will enable 
to populate a mass/luminosity space as a function of time to calibrate models of planetary formation and
evolution. This will soon come using the synergy between observing
techniques, such as radial velocity, astrometry with \textit{GAIA} and
DI that will overlap with the new generation of planet imagers SPHERE
and GPI or more likely the the ones that will be offered on the ELTs.

%%%%%%%%%%%%%%%%%%%%%%%%%%%%%%%%%%%%%%%%%
\subsection{Physics of imaged planets} 
\label{subsec:physics}

\begin{figure}[t]
\begin{center}
\includegraphics[height=7cm]{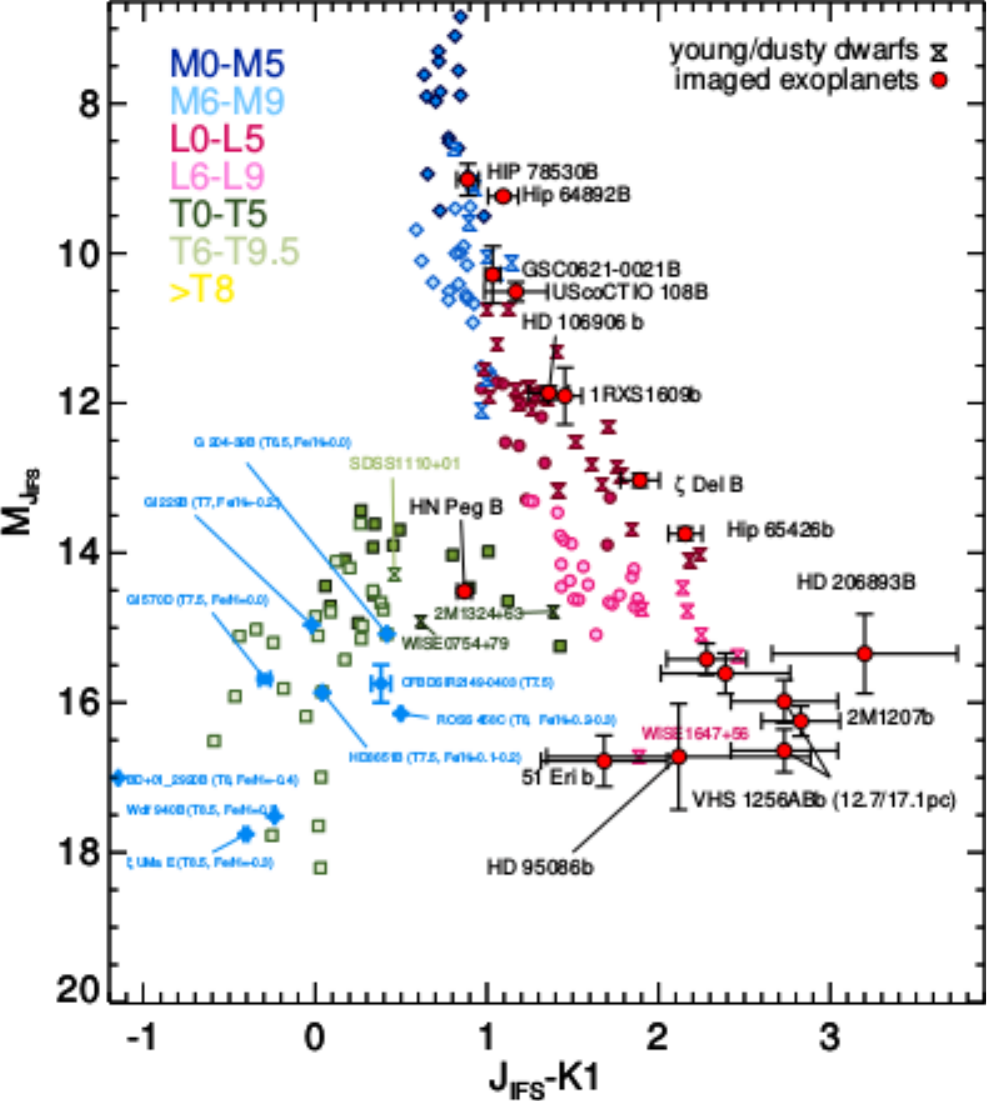} \hspace{0.5cm}
\includegraphics[height=7cm]{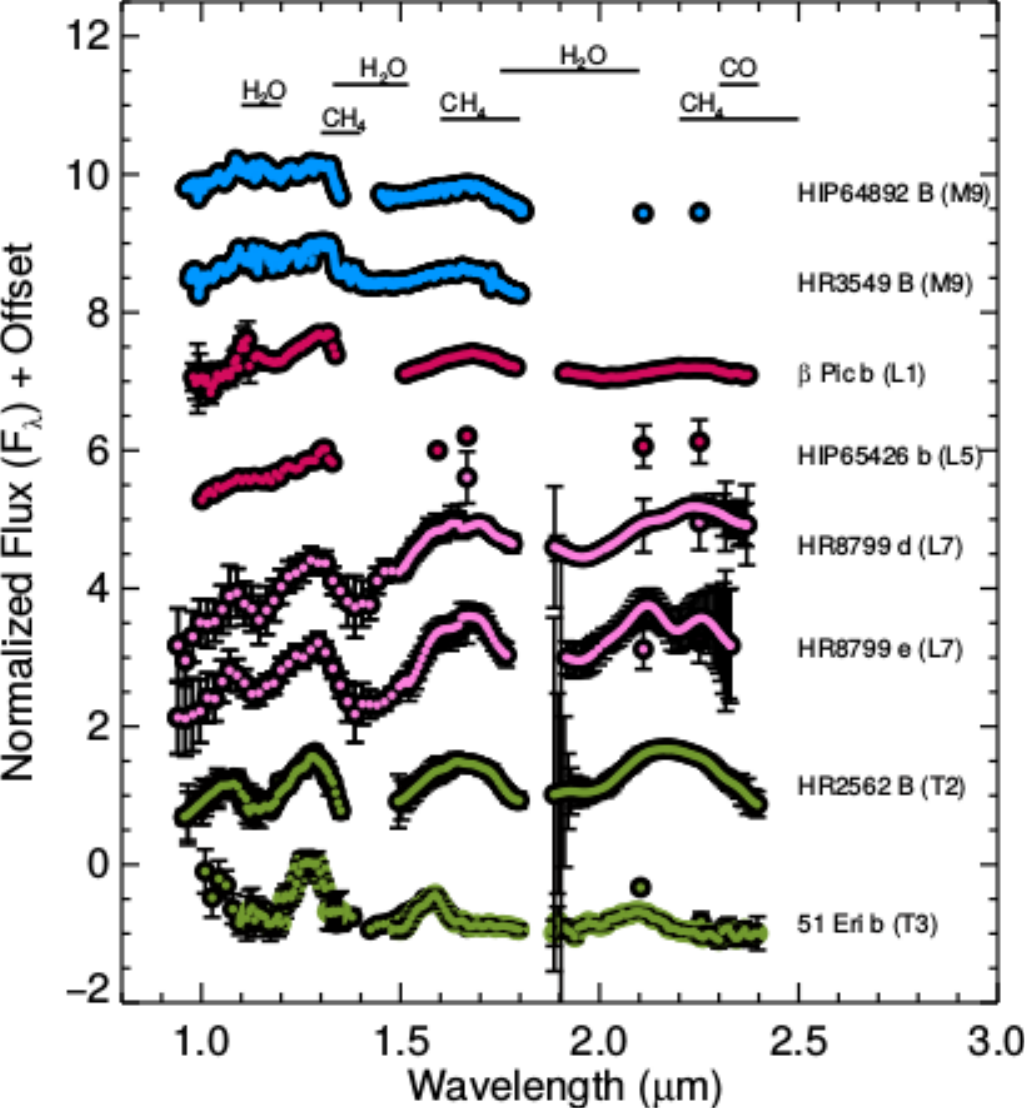} 

\caption{\textit{Left:} color-magnitude diagrams comparing the young imaged exoplanets with field M, L, and T dwarfs with known parallax
  measurements. \textit{Right:} Sequence of exoplanet's spectra characterized in the course of the SHINE and GPIES surveys\cite{mesa2016,chilcote2015,derosa2016,macintosh2015,samland2017,mesa2018,cheetham2018,greenbaum2018,chauvin2017b,konopacky2016,samland2017,bonnefoy2017}.}

\label{fig:colors}
\end{center}
\end{figure}

Today, most imaged exoplanets are young, late-M, L to early T-types
exoplanets with dusty atmospheres (see Table~\ref{tab:planets}). The
majority occupies a slightly distinct space compared to field brown
dwarfs in near-IR color-magnitude diagrams. They are redder than field
brown dwarfs, their older counterparts. For some cases like 2M1207\,b,
HR8799\,b and VHS\,1256\,b, they are also underluminous (see
Fig.~\ref{fig:colors}, \textit{Left}). The most plausible explanation is the presence
of unusual thick clouds due to low surface gravity conditions in
these young planetary atmospheres. Modification of the cloud thickness
and distribution is corroborated by recent variability studies that
show higher-amplitude variability in young L dwarfs than old ones
\cite{metchev2015}. In that perspective, future high-precision
photometric monitoring are very promising to probe time-variability in
the planet's flux and/or spectra and explore the physical and
dynamical processes at play in young exoplanetary atmospheres. To 
further characterize the physical properties of imaged planets
(luminosity, effective temperature, surface gravity, radius and
ultimately the mass), a logical approach is to use multi-wavelength
photometry to constrain the exoplanet's spectral energy distribution
(SED). All imaged planets have good parallaxes measurements thanks to
the primary star. Planetary SEDs of current imaged planets peak
between 1.2 to 5.0~$\mu$m for $T_{\rm{\rm{eff}}}\,=\,2400$~K to
$T_{\rm{eff}}=600$~K, respectively. The use of bolometric correction
calibrated with observations of field brown dwarfs is rather risky as
small variation in the cloud and dust
properties may actually affect the flux re-distribution over the SED
and therefore biases the luminosity determination. The challenge is now to extend 
the spectral coverage to lift degeneracies in the fitting of the 
physical and atmospheric properties, and \textit{JWST} will be very precious in that sense.
For SED characterization, the case of $\beta$ Pic\,b is nowadays the most favorable one as the
planet is the brightest one imaged to date with an apparent magnitude of 
$H = 13.5 \pm0.2$~mag (HR\,8799\,bcde have $H$-band magnitudes of
17.7, 16.8, 16.8, 16.7~mag, respectively). Based on a combination of MagAO, NaCo, NICI 
observations from 0.9 to 5.0$~\mu$m, accurate bolometric luminosity of $log(L/L_{\odot}) = −3.78\pm0.03$ could be derived 
\cite{morzinski2015}. For an age of $23\pm3$\,Myr, it corresponds to predicted hot-start 
physical parameters of $T_{\rm{eff}}=1708\pm23$~K, radius of $R = 1.45\pm0.02~R_{\rm{Jup}}$, 
and mass of $M = 12.7\pm0.3~M_{\rm{Jup}}$, compatible with previous atmosphere model fitting 
in near-IR \cite{bonnefoy2013} and current radial velocity constraints for the planet dynamical mass\cite{lagrange2012b}.

For further insight on the atmospheric properties of the direct imaged planets, spectroscopic observational at higher spectral resolution are particularly interesting to study the atmosphere composition through the various atomic and molecular lines, to explore  the impact  of low-gravity conditions, to confirm the existence of clouds and constrain their micro-physics (grain size distribution and composition), finally to test the effect of alternative non-equilibrium processes at play. As proxies, the use of young brown dwarfs as
exoplanets analogs has proven to be extremely useful to test the reliability of
atmospheric models as we cannot currently distinguish both classes of
objects. It enabled to confirm for instance that low-gravity conditions 
are responsible for the temperature shift by $200-500$~K observed in the 
classical $T_{\rm{eff}}$--SpT 
classification between young and field dwarf, the inhibition of the formation of methane 
at the L/T transition that probably occurs at lower effective temperatures or the enhanced production of 
dust forming thick clouds and potentially responsible for increased photometric variabitilies\cite{metchev2015}. These results can directly be compared to the building sequence of low-resolution spectra of imaged exoplanets, now regularly acquired with the new generation of planet imagers, to further understand the physics of their atmospheres and the influence of effective temperature, low-gravity conditions or clouds in shaping their spectral morphology (see Fig.\,\ref{fig:colors}, \textit{Right}). With the increased of ($R_\lambda > 2000$) spectral resolution,  the measurement
of molecular abundances such as water, carbon-dioxide and monoxide or
methane becomes possible as done for HR\,8799\,b \cite{barman2015}. It may give a first
estimation of the carbon-to-oxygen ratio in the planet atmosphere
(0.6-0.7 for HR\,8799\,b). When compared to the stellar C/O ratio, it
can indicate some enrichment by heavy elements as for the giant planets
of our Solar systems relative to our Sun ($\sim4$ for Jupiter, $\sim7$
for Saturn and $\sim45$ for Uranus and Neptune). However, despite
non-negligible errors in the determination of C/O ratio using that
technique, a direct link between C/O ratio and formation mechanisms
by core accretion or gravitational instability is not trivial as 
additional physical processes might also enrich the planetary envelope 
after the giant planet formation.

%%%%%%%%%%%%%%%%%%%%%%%%%%%%%%%%%%%%%%%%%
\subsection{Orbits \& planetary architectures}

Among the known imaged planets which are mostly located at wide orbits
($\ge10$~au), orbital motion has been resolved for a very few cases. It
actually concerns the case of $\beta$\,Pic\,b, Fomalhaut\,b,
HR8799\,bcde, HD\,95086\,b, 51\,Eri\,b, and GJ504\,b 
and $\kappa$ And\,b. Vast efforts have been devoted to monitor the two emblematic cases $\beta$~Pic and HR\,8799 for more than a decade now. Since the first epoch discovery of $\beta$\,Pic\,b in November 2003 with NaCo at VLT, various ground-based instruments have been used and enabled to follow more than half of the orbit around its stellar host. For the four planets around HR\,8799, the longest baseline has been achieved for the b planet thanks to the re-analysis of archived \textit{HST/NICMOS} data from 1998. Although accurate astrometric monitoring with deep imagers is often a meticulous task to achieve due to the relatively small field of view of these instruments that could lead to astrometric calibration systematics, we see now that the 1-2\,mas astrometric precision achieved by current planet imagers with adequate calibration strategy allows to study, in addition to the planet's orbit, the global system architecture including the planet-planet and planet-disk interactions and the system stability. The orbital monitoring of $\beta$~Pic\,b enabled for instance to witness
the planet's revolution around its star, passing behind the star from
a North-East location to South-West, reaching quadrature in Summer
2013 and then coming back toward us for a possible transit in mid-
2017 or early-2018 to be confirmed as soon as the planet is recovered 
on the North-East location. In that specific case, sophisticated 
statistical approaches using the Markov-chain Monte Carlo (MCMC)
Bayesian analysis technique have been used as they offer the advantage 
to be less sensitive to the initial conditions assumed to
derive and systematically explore the parameter space of orbital
parameters. In the case of $\beta$~Pic\,b, the results indicate a semi-major axis
distribution peaking at 8--10\,au with most eccentricities by
$e\le0.20$, and extremely concentrated inclined solutions 
\cite{chauvin2012a,wang2016}.  Similar studies have been conducted for 51\,Eri\,b and the young solar analogs HR\,8799\,bcde 
planets\cite{pueyo2015,maire2015,zurlo2016,konopacky2016} and HD\,95086\,b \cite{rameau2016,chauvin2018}. 
An interesting question addressed by DI in the context of the multiple planetary 
system HR\,8799 is connected to the system's dynamical stability. The
long-term astrometric monitoring actually enabled to characterize the planet
orbits, to check for coplanarity, and ultimately to test the dynamical
stability of the whole system. The derived orbital
properties of each planet showed that all planets evolve
on coplanar and circular orbits except HR\,8799\,d which seems to be
misaligned by 15--20 degrees and significantly more eccentric
($\sim0.3$). This configuration likely suggests remaining imprints of
dynamical interactions between the planets in this system. The system
seems however stable as indicated by the existence of several mean
motion resonances: a period ratio between 5:2 or 3:1 between b-c, 2:1
between c-d, and 3:2 or 2:1 between d-e. Even if HR\,8799\,d most
likely does not orbit in the same plane as HR 8799c and e, the period ratios 
involving these planets do
not rule out a 1e:2c:4d Laplace resonance\cite{pueyo2015}.

\begin{figure}[t]
\begin{center}
\includegraphics[height=7cm]{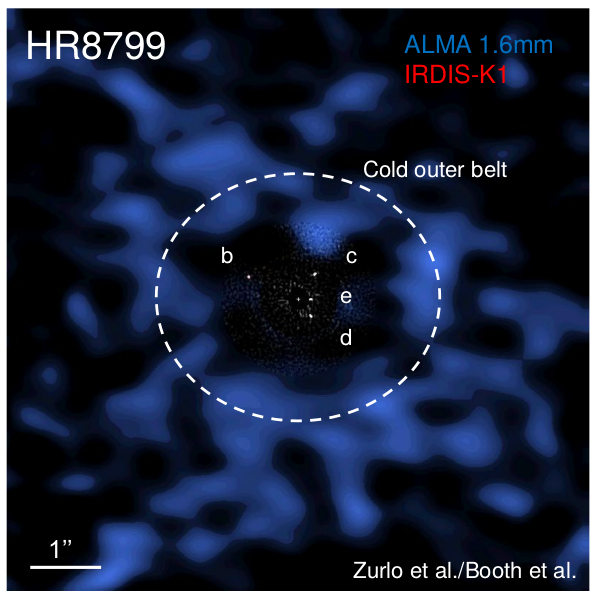}  \hspace{0.5cm}
\includegraphics[height=7cm]{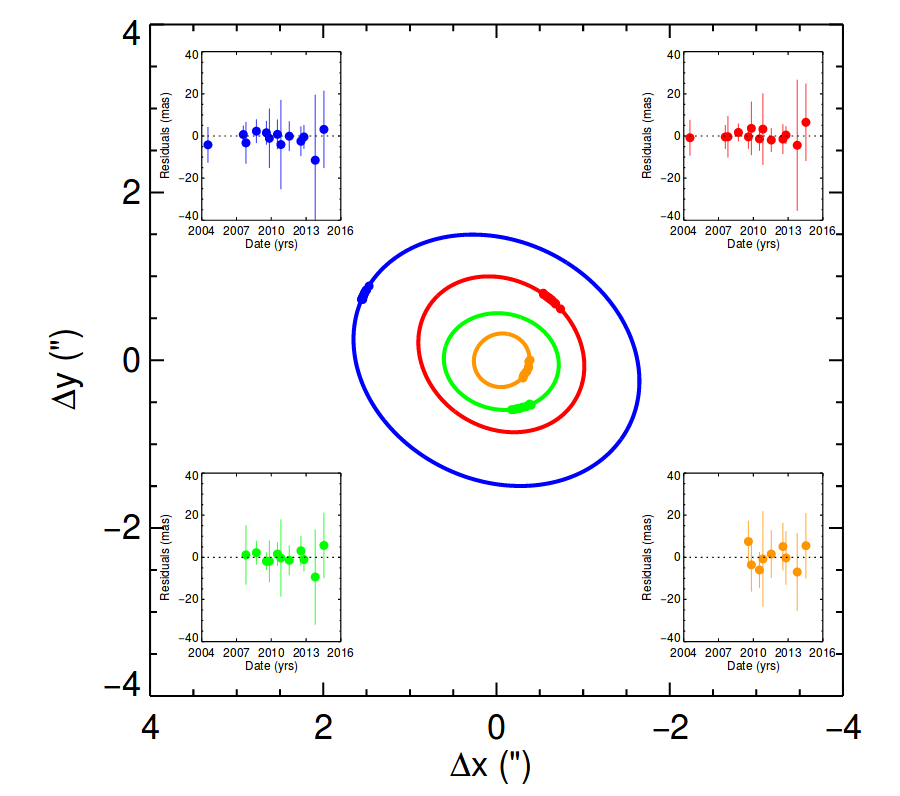}
\caption{\textit{Left:} Composite map showing the ALMA 1.3mm emission around HR\,8799 and revealing the faint outer belt located at more than 100\,au\cite{booth2016} together with the SPHERE/IRDIS observations of the four planets bcde\cite{zurlo2016}.
\textit{Right:} Low-eccentricity orbits that are consistent with the compiled astrometry for the four planets\cite{konopacky2016}. The side panels shown the size of the residuals to each of these fits.}
\label{fig:orbit}
\end{center}
\end{figure}

Regarding the global architecture and the circumstellar environment, scattered-light
observations bring additional constraints and accurately map the
morphology of the proto-planetary or debris disk generally discovered before the planet discoveries.  For
$\beta$~Pic, the disk observations revealed mainly a nearly edge-on
system composed of a main disk observed beyond 80~au with a position
angle of $\sim209.0^o$, and an inner warped component (at less than 80\,au), 
inclined by $2-5^o$ with respect to the main disk position angle
(i.e with a position angle of $\sim211.0-214.0^o$. Dynamical simulations\cite{mouillet1997,augereau2001,dawson2011} demonstrated that the presence of a planet
orbiting the star at 10\,au, misaligned with the main disk, could
actually form and sustain the $\beta$\,Pic inner warped disk. The
current distribution of longitude of ascending node $\Omega$ 
found for $\beta$~Pic\,b, when compared with the main disk and the warp orientations,
supports an orbital plane for $\beta$~Pic\,b that does not coincide
with the main disk midplane, but more probably with the warp
component. The orbital solutions therefore confirm that the planet is
the perturber that excites the disk of planetesimals, forcing them to
precess about its misaligned orbit. Simultaneous observations of the
planet and disk position unambiguously confirm this conclusion \cite{lagrange2012a}. Similar studies have been conducted for the young solar analogs HR\,8799\cite{pueyo2015,maire2015} and HD\,95086\cite{rameau2016,chauvin2018}. Here again, 
the planet location(s) can be compared to
the dust spatial distribution and architecture. At ages $\ge10$~Myr,
the gas is expected to have dispersed, giant planets to have formed
and the dynamics of planetesimals to be influenced by the presence of
giant planets. For both cases, the orbital properties of the imaged
planets suggest that the planets are shepherded by an inner warm belt
and an outer cold belt revealed by a two-temperature component model
fitting their observed spectral energy distributions. This
configuration is very similar to the one of our solar system with the
asteroid and the Kuiper belts separated by giant planets. Their origin
remains unclear. We could be witnessing two independent belts
undergoing normal collisional evolution \cite{kennedy2014}. The
inner components may also be linked to the outer belt via inward
scattering of material by intervening planets \cite{bonsor2012}. The second generation of planet imagers, by discovering new exoplanetary systems,
should significantly extend this list in the coming years to study the
orbital diversity of theses systems. It already started with the GPI and SPHERE discoveries around 51\,Eri\cite{macintosh2015}, HIP\,65426\cite{chauvin2017} and more recently PDS\,70\,b, but also of the interesting young brown dwarf companions around HR\,2562\cite{konopacky2016} and HIP\,64892\cite{cheetham2018}. Further discoveries of exoplanetary systems in DI and in
combination with other techniques should enable to more systematically
study the diversity of the planetary architectures at young
ages. Young, nearby associations with ages spanning from a few
Myrs to a hundred offer an ideal laboratory to explore various stages
of evolution of planetary systems from the giant planet formation
phase to more catastrophic events like the heavy bombardment of our
Solar system caused by Jupiter and Saturn migration.

%%%%%%%%%%%%%%%%%%%%%%%%%%%%%%%%%%%%%%%%%
\subsection{Occurrence \& formation} 

With the manna of exoplanet discoveries since the 51 Peg\,b announcement \cite{mayor1995}, the diversity of systems found (Hot Jupiters, irradiated and evaporating planets, misaligned planets with stellar spin, circumstellar and circunbinary planets in binaries, telluric planets in habitable zone, discovery of Mars-size planet...), the theories of planetary formation have drastically evolved to digest these observing constraints. However, we are still missing the full picture and some key fundamental questions still lack answers like: the physical processes at play to pass the km-size barrier to form planetary cores, the physics of accretion to form
planetary atmospheres, the formation mechanisms to explain the existence of giant planets at wide orbits, the physical properties of young Jupiters, the impact of planet-planet and planet-disk interaction in the final planetary system architecture, the influence of the stellar mass and stellar environment in the planetary formation processes. Neither core accretion plus gas capture\cite{pollack1996} (CA) nor disk fragmentation driven by gravitational instabilities\cite{cameron1978} (GI) can globally explain all current observable from planet hunting techniques. Alternative mechanisms are then proposed, such as pebbles accretion\cite{lambrechts2012} to enable CA to operate at wide orbits, inward/outward migration or planet-planet\cite{crida2009} or simply the possibility to have several mechanisms forming giant planets operating in concert\cite{boley2009}.A stellar-like mechanism like gravo-turbulent fragmentation could be also
an alternative solution to form massive planetary mass companions at
wider separations ($\ge5-10$~au) in the earliest phase of the disk's
lifetime. In this context, one final key element addressed by DI surveys concerns the occurrence of giant planets beyond 5--10~au to actually test these theories. The rate of discovery in DI is currently low despite the large number of stars
observed. These non-detection can however be used to actually
derive the survey completeness and the frequency of the giant planet population. 

\begin{figure}[t]
\begin{center}
\includegraphics[height=5cm]{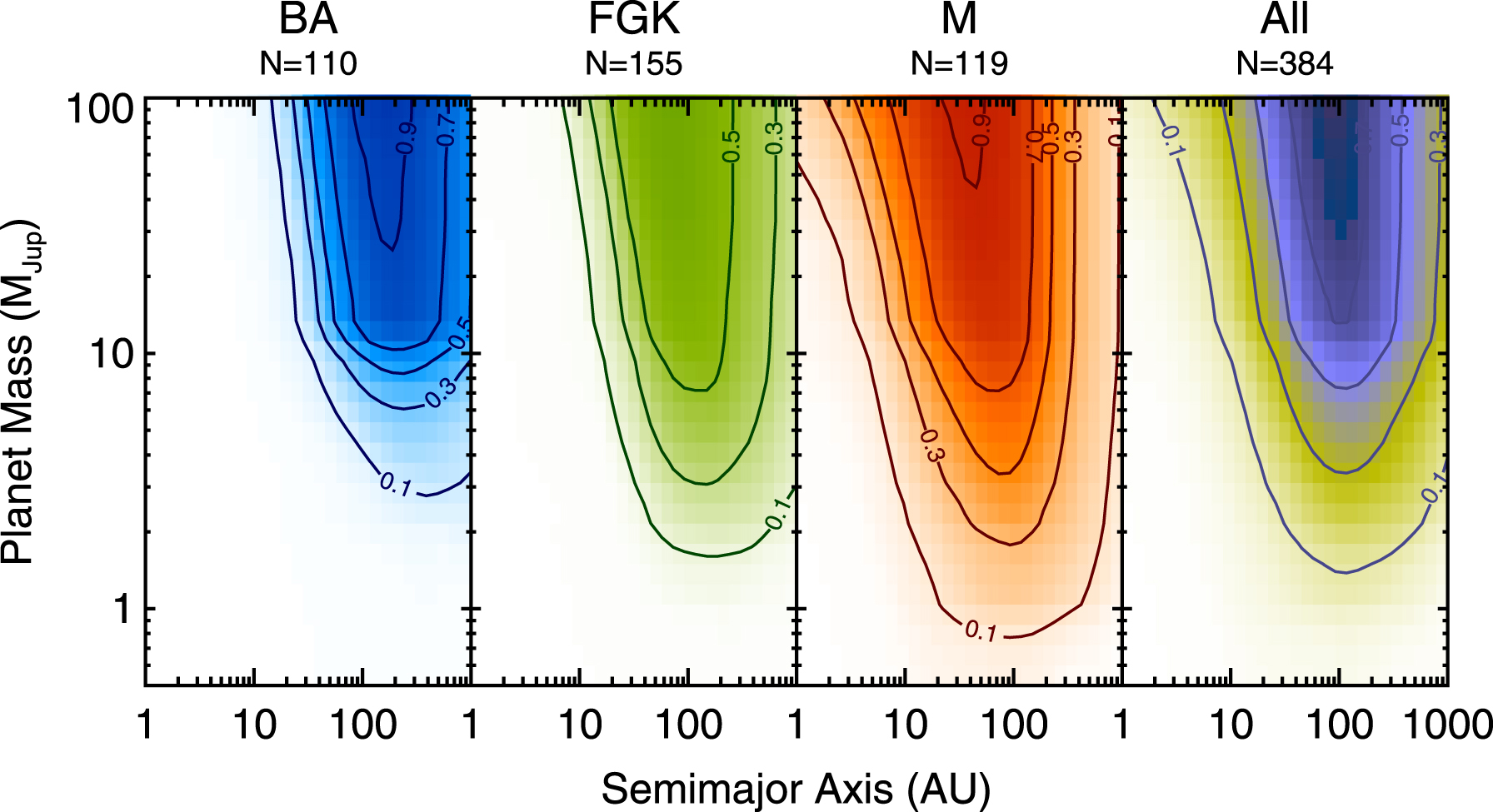} \hspace{0.1cm} \includegraphics[height=4.8cm]{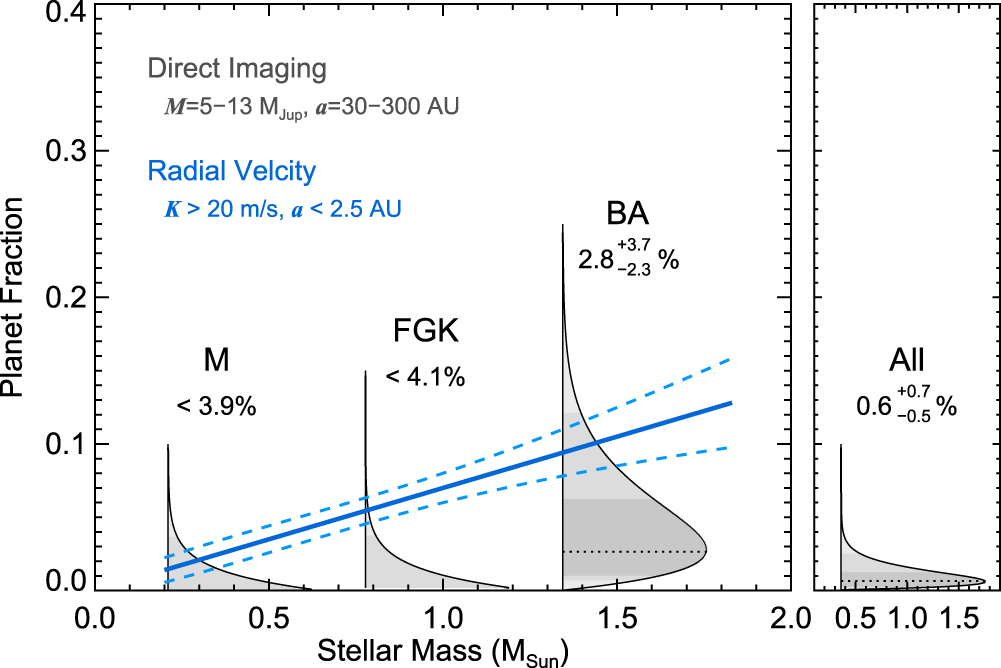} 
\caption{\textit{Left:} Mean sensitivity maps from a meta-analysis of 384 unique stars with published high-contrast imaging observations\cite{bowler2016}. M dwarfs provide the highest sensitivities to lower planet masses in the contrast-limited regime. Altogether, current surveys probe the lowest masses at separations of ~30–300 au. Contours denote 10\%, 30\%, 50\%, 70\%, and 90\% sensitivity limits. \textit{Right:} Probability distributions for the occurrence rate of giant planets from a meta-analysis of DI surveys in the literature.}
\label{fig:stat}
\end{center}
\end{figure}

Early studies\cite{lafreniere2007,nielsen2008,chauvin2010} have developed statistical analysis tools to exploit the complete deep imaging performances and derive the detection probabilities of their surveys by simulating  synthetic planet population described by various sets of (mass, eccentricity, semi-major axes) parametric distributions confronted to the survey detection limits. They show that the first generation of surveys were mostly sensitive to giant planets more massive than $5\,M_{\rm{Jup}}$ for semi-major axis between typically 30 to 300\,au (see Fig.\,\ref{fig:stat}, \textit{Left}).
Pushing that logic, the recent large meta-analysis\cite{bowler2016} of 
384 unique and single young (5--300 Myr) stars spanning stellar masses between 0.1 and $3.0\,M_{\odot}$ and combining several DI surveys illustrates that the current overall occurrence rate of $5-13M_{\rm{Jup}}$ companions at orbital distances of 30--300 au is ${0.6}_{-0.5}^{+0.7}\% $ (assuming hot-start evolutionary models). Splitting the analysis for separate bins of masses containing BA stars (110 targets), FGK stars (155 targets), and M dwarfs (119 targets), it shows that the frequency of planets orbiting BA, FGK, and M stars is ${2.8}_{-2.3}^{+3.7}\%$, $<4.1\%$, and $<3.9\%$, respectively (see Fig.\,\ref{fig:stat}, \textit{Right}). Although there are hints of a higher occurrence rate of giant planet around massive stars analogous to the well-established correlation at small separations, the trend is not yet statistically significant at wide orbital distances. It will require larger sample sizes in each stellar mass bin to unambiguously test this correlation, which is supporting the need for systematic DI surveys like the ones currently carried out by the GPI and SPHERE large programs of more than 600 young, nearby stars of various masses and ages.

\begin{figure}[t]
\begin{center}
\includegraphics[height=7cm]{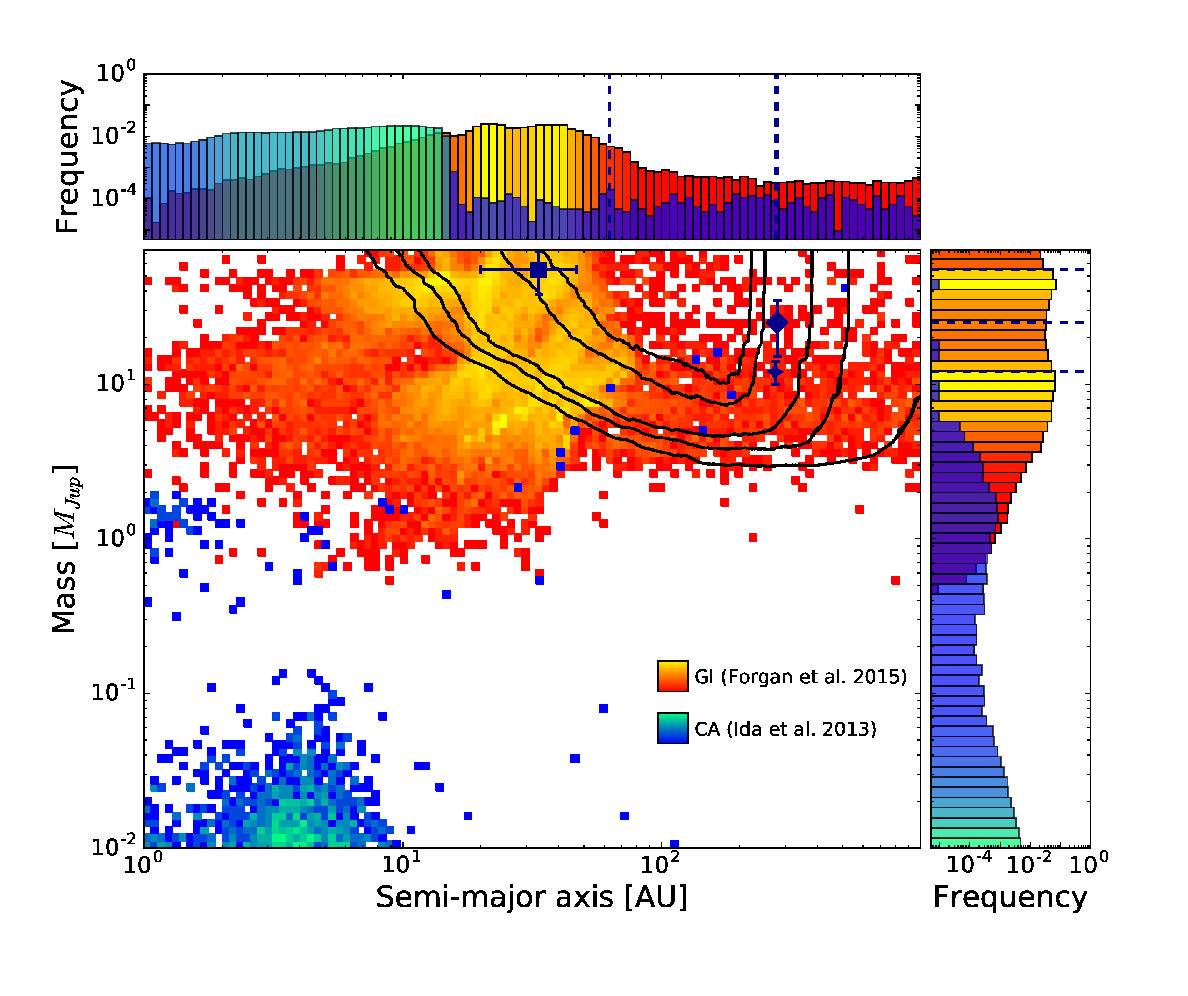} 
\caption{Density plots representing scattered populations based on GI \cite{forgan2015} and CA \cite{ida2013} compared to the detections in our sample and to the mean detection probabilities of the observations (contour lines are at 5\%, 25\%, 50\%, 90\%, and 95\%)\cite{vigan2017}. Density colors go from dark blue (low occurrence) to cyan (high occurrence) for the CA population, and from red (low occurrence) to yellow (high occurrence) for the GI population. The histograms on top and on the right represent the relative frequency in each bin of semi-major axis and planetary mass respectively. The histograms take into account the whole population, including the planets that are outside of the visibility window of the main plot. The semi-major axis and mass of the known companions are represented with dashed lines in the histograms.}
\label{fig:formation}
\end{center}
\end{figure}

In addition to the determination of the occurrence of giant planets at wide orbits, a further step in such a statistical exploitation of the survey completeness is a direct comparison with the the predictions of planetary formation models\cite{janson2011,rameau2013a}. Here again, Monte-Carlo simulations can be used to compare the sensitivity of DI surveys with the state-of-the-art population synthesis models. This is actually illustrated with Fig.\,\ref{fig:formation} that shows two populations of giant planets predicted by the CA scenario and the GI one. Both models include scattering effects between planetary embryos. They are compared with the completeness of the NaCo-LP survey\cite{chauvin2015,vigan2017} of about 200 young, nearby solar-type stars published together with the Direct Imaging Virtual Archive\footnote{DIVA, http://cesam.lam.fr/diva/}. This comparison highlights the fact that the first geenration of DI surveys marginally explored the bulk of giant planets predicted by CA, that planets formed by GI are rare considering that they should be easily detectable in most current DI surveys, finally that, although GI is not common, it predicts a mass distribution of wide-orbit massive companions closer to what is observed compared to current CA predictions. In a near future, the synergy of all observing techniques will be very precious to constrain the occurrence rate of giant planets at all separations, but more importantly the properties of their underlying mass, separation, and
eccentricity distributions to shed more light on the formation and evolution mechanisms of giant planets.

% Planet Formation
% Conclusions

% Another way is also to Comparison to models of formation (Janson, Rameau, Vigan et al.)
% Global view at all orbits; importance of synergies of techniques

%Once the
%detection limit of the survey derived, one needs
%to correct for the projection effect of the observations. 

%%%%%%%%%%%%%%%%%%%%%%%%%%%%%%%%%%%%%%%%%%%%%%%%%%%%%%%%%%%%%%%%%%%%%%%%%%%%%
\section{Conclusions \& perspectives}
\label{sec:conclu}  % \label{} allows reference to this section

Today's success of DI relies on a sophisticated instrumentation designed to
detect a faint planetary signal, angularly close to a bright host star. It also relies on a fine target selection of young, nearby stars sharing common kinematics, photometric and spectroscopic properties. This combination enabled the discovery of the first exoplanets and/or planetary mass companions at large
physical separations ($>100$\,au) or with small mass ratio with their primaries. This success was followed by breakthrough discoveries of closer and/or lighter exoplanets. Each one of these discoveries has proven to be rich in terms of scientific exploitation and characterization to directly probe the presence of planets in their birth environment, to explore the orbital, physical and spectral characterization of young Jupiters, or more globally explore the young planetary system architectures. Vast efforts are now devoted to systematic searches of exoplanets in DI with an increasing number of large scale surveys. With bigger samples and enhanced detection performances, new large surveys of 100+ nights with the second generation of planet finders will offer unprecedented statistical constraints on the occurrence of giant planets at wide orbits. With the rich perspective of new/upgraded instruments from the ground (VLT/ESPRESSO, CFHT/SPIROU, ESO3.6/NIRPS, LBT/iLocater, SCExAO/CHARIS, Keck/KPIC, VLT/ERIS, LBT/SHARK(S), MagAO-X, Gemini/GPI+, VLT/SPHERE+...) and space missions (\textit{GAIA, TESS, CHEOPS, JWST, PLATO, WFIRST, ARIEL...}), devoted to the study of exoplanets, we can hope to obtain a complete census of nearby planetary systems within a decade from now. This will end the era of exoplanet surveys and will open a characterization phase for the extremely large telescopes (ELT, TMT, GMT) or the dedicated space missions from space (LUVOIR, HABEX...) that might exploit the synergy of various
planet hunting techniques including DI with the ultimate objective to detect and characterize the first bio-signatures at the horizon 2030--2040.

%\begin{figure}[t]
%\begin{center}
%\includegraphics[height=6cm]{David_exoplanet_images.pdf} 
%\caption{Non-exhaustive family portrait of young exoplanets and planetary mass companions discovered in DI.}
%\label{fig:discoveries}
%\end{center}
%\end{figure}

\acknowledgments % equivalent to \section*{ACKNOWLEDGMENTS}      
This review is the fruit of the successful work of the large direct imaging community over the years. For the preparation of this review, I would like to particularly thank D. Mouillet, M. Bonnefoy, L. Close, C. Mordasini, M. Booth, P. Kern, L. Pueyo, J. Rameau and A. Vigan for their help.

%This unnumbered section is used to identify those who have aided the authors 
%in understanding or accomplishing the work presented and to acknowledge 
%sources of funding.  

% References
\bibliography{report} % bibliography data in report.bib
\bibliographystyle{spiebib} % makes bibtex use spiebib.bst

\end{document}